\documentclass[10pt,preprint]{aastex}

\shorttitle{SN 2006X in M100} \shortauthors{Wang et al.}

\def\gsim{\;\lower4pt\hbox{${\buildrel\displaystyle >\over\sim}$}\;}
\def\lsim{\;\lower4pt\hbox{${\buildrel\displaystyle <\over\sim}$}\;}
\def\grls{\;\lower4pt\hbox{${\buildrel\displaystyle >\over <}$}\;}

\begin{document}

\title{Optical and Near-Infrared Observations of the Highly
Reddened, \\ Rapidly Expanding Type Ia Supernova 2006X in M100}

\author{Xiaofeng Wang\altaffilmark{1,2}, Weidong Li\altaffilmark{1},
Alexei V. Filippenko\altaffilmark{1}, Kevin
Krisciunas\altaffilmark{3}, Nicholas B. Suntzeff\altaffilmark{3}, \\
Junzheng Li\altaffilmark{2}, Tianmeng Zhang\altaffilmark{4},
Jingsong Deng\altaffilmark{4}, Ryan J. Foley\altaffilmark{1}, Mohan
Ganeshalingam\altaffilmark{1}, \\ Tipei Li\altaffilmark{2}, YuQing
Lou\altaffilmark{2}, Yulei Qiu\altaffilmark{4}, Rencheng
Shang\altaffilmark{2}, Jeffrey M. Silverman\altaffilmark{1}, \\
Shuangnan Zhang\altaffilmark{2}, and Youhong Zhang\altaffilmark{2}}

\altaffiltext{1}{Department of Astronomy, University of California,
Berkeley, CA 94720-3411, USA; wangxf@astro.berkeley.edu .}
\altaffiltext{2}{Physics Department and Tsinghua Center for
Astrophysics (THCA), Tsinghua University, Beijing, 100084, China;
wang\_xf@mail.tsinghua.edu.cn .} \altaffiltext{3}{Physics
Department, Texas A\&M University, College Station, Texas, 77843 }
\altaffiltext{4}{National Astronomical Observatories of China,
Chinese Academy of Sciences, A20, Datun Road, Beijing 100012,
China.}

\begin{abstract}

We present extensive optical ($UBVRI$), near-infrared ($JK$) light
curves and optical spectroscopy of the Type Ia supernova (SN) 2006X
in the nearby galaxy NGC 4321 (M100). Our observations suggest that
either SN 2006X has an intrinsically peculiar color evolution, or it
is highly reddened [$E(B - V)_{host} = 1.42 \pm 0.04$ mag] with $R_V
= 1.48 \pm 0.06$, much lower than the canonical value of 3.1 for the
average Galactic dust. SN 2006X also has one of the highest
expansion velocities ever published for a SN~Ia. Compared with the
other SNe~Ia we analyzed, SN 2006X has a broader light curve in the
$U$ band, a more prominent bump/shoulder feature in the $V$ and $R$
bands, a more pronounced secondary maximum in the $I$ and
near-infrared bands, and a remarkably smaller late-time decline rate
in the $B$ band. The $B-V$ color evolution shows an obvious
deviation from the Lira-Phillips relation at 1 to 3 months after
maximum brightness. At early times, optical spectra of SN 2006X
displayed strong, high-velocity features of both intermediate-mass
elements (Si, Ca, and S) and iron-peak elements, while at late times
they showed a relatively blue continuum, consistent with the blue
$U-B$ and $B-V$ colors at similar epochs. A light echo and/or the
interaction of the SN ejecta and its circumstellar material may
provide a plausible explanation for its late-time photometric and
spectroscopic behavior. Using the Cepheid distance of M100, we
derive a Hubble constant of $72.8 \pm 8.2$ km s$^{-1}$ Mpc$^{-1}$
(statistical) from the normalized dereddened luminosity of SN 2006X.
We briefly discuss whether abnormal dust is a universal signature
for all SNe~Ia, and whether the most rapidly expanding objects form
a subclass with distinct photometric and spectroscopic properties.

\end{abstract}


\section{Introduction}

Type Ia supernovae (SNe~Ia) are arguably the most accurate tools
available for probing the expansion history of the Universe, as
their luminosities at maximum light can be calibrated to an
uncertainty of $\sim$15\% via several empirical relations between
their luminosity and light/color curves (Wang et al. 2006, and
references therein). Based on studies of SN~Ia apparent brightness
as a function of redshift, Riess et al. (1998) and Perlmutter et al.
(1999) were the first to propose that the expansion of the Universe
is currently accelerating (for a review see, e.g., Filippenko 2005).
This remarkable result, which suggests that the Universe is
primarily composed of mysterious dark energy, was subsequently
confirmed by more studies using SNe~Ia (Barris et al. 2003; Tonry et
al. 2003; Knop et al. 2003; Riess et al. 2004, 2007; Astier et al.
2006; Wood-Vasey et al. 2007), and by independent methods such as
the power spectrum of fluctuations in the cosmic microwave
background radiation (e.g., Spergel et al. 2003, 2007) and baryon
acoustic oscillations (Eisenstein et al. 2005).

Despite the success of using SNe~Ia as cosmological probes, a
disconcerting fact is that details of the properties of SN~Ia
progenitors and their environments remain poorly understood. It is
generally believed that a SN~Ia is produced by a carbon-oxygen
(C$-$O) white dwarf (WD) accreting matter from a companion star in a
binary system. However, the nature of the companion star in the
binary system is still unclear: it could be a main-sequence star, a
subgiant, or an evolved red giant (Nomoto, Iwamoto, Koshimoto 1997),
or even another degenerate C$-$O WD (Iben \& Tutukov 1984; Webbink
1985).

Studying the absorption by dust toward SNe~Ia may provide a unique
way to peer into their local environments and hence set constraints
on the properties of their progenitor system. If the dust enveloping
the SNe is totally or partially produced by the progenitor
evolution, the ratio of the total to selective absorption [$R_V =
A_V/E(B-V)$] may differ from that of normal interstellar dust (e.g.,
Wang L 2005). Moreover, accurate determination of the absorption
toward a SN~Ia helps establish the accurate luminosity required for
precision cosmology.

Several studies of global color fits to SNe~Ia have suggested that
the dust obscuring SNe~Ia in their host galaxies may differ from the
dust observed in the Milky Way Galaxy (Riess et al. 1996; Phillips
et al. 1999; Altavilla et al. 2004; Reindl et al. 2005; Wang et al.
2006): the measured value of $R_{V}$ is often found to be smaller
than the canonical Galactic value of 3.1. We measure an average
$R_{V} = 2.4 \pm 0.2$ from a compilation of published $R_{V}$
values\footnote{Methods that do not separate the intrinsic color
variation and the host-galaxy reddening from the observed color
would give even lower $R$ values (Tripp 1997; Astier et al. 2006;
Wang L et al. 2006a).}. The sample of well-observed, highly
extinguished SNe~Ia is small but growing, offering the opportunity
to study variations of the extinction law on an object-by-object
basis. In almost all cases, $R_{V}$ is found to be smaller than 3.1
(Krisciunas et al. 2006a; Elias-Rosa et al. 2006, hereafter ER06).
It is interesting to investigate whether $R_{V}$ smaller than 3.1 is
universal to all SNe~Ia, or just applies to a few or highly reddened
events.

SN 2006X is a bright, nearby, highly reddened SN~Ia, and is
favorable for a detailed study of reddening and dust properties. The
recent report of variable, interstellar absorption Na~I~D lines the
in high-resolution spectra of SN 2006X indicates of the presence of
circumstellar material (CSM) around SN 2006X, which may be related
to the stellar wind blown from the companion star (Patat et al.
2007). Here, we present extensive photometric and spectroscopic
observations of SN 2006X, enabling us to study its intervening dust
and other characteristics. The paper is organized as follows. The
observations and data reduction are described in \S 2, while \S 3
presents the $UBVRIJK$ light curves, the color curves, and the
extinction estimate. The spectral evolution is given in \S 4. We
determine the distance and luminosity of the SN in \S 5. Our
discussion and conclusions are given in \S 6.

\section{Observations and Data Reduction}

SN 2006X was discovered independently on 7.10 February 2006 (UT
dates are used throughout this paper) by S. Suzuki and M. Migliardi
(IAUC 8667), with J2000 coordinates $\alpha$ =
12$^{h}$22$^{m}$53$^{s}$.90 and $\delta$ =
15$^{\circ}$48$^{'}$32$^{''}$.90. It is $12''$ W and $48''$ S of the
center of the nearby galaxy NGC 4321 (M100), a grand-design Sbc
galaxy in the Virgo cluster having a Cepheid distance of $30.91 \pm
0.14$ mag (Freedman et al. 2001). A spectrum of SN 2006X taken
shortly after its discovery shows it to be an early SN~Ia (Quimby et
al. 2006), similar to that of SN 2002bo 1--2 weeks before maximum
light (Benetti et al. 2004) but with a redder continuum.

\subsection{Photometry}

Our photometric observations of SN 2006X began on 2006 February
8.15, $\sim$11 days before $B$-band maximum. Data were obtained with
three telescopes: the 0.80~m THCA-NAOC Telescope (TNT) at Beijing
Xinglong Observatory (BAO) in China, the 0.76~m Katzman Automatic
Imaging Telescope (KAIT; Filippenko et al. 2001) at Lick Observatory
in the U.S., and the 1.3~m telescope at Cerro Tololo Inter-American
Observatory (CTIO) in Chile. The TNT observations were obtained
using a $1340 \times 1300$ pixel back-illuminated charge-coupled
device (CCD) with a field of view (FOV) of $11.5' \times 11.2'$
($0.52''$ pixel$^{-1}$), the KAIT observations were performed using
an Apogee AP7 camera with a FOV of $6.6' \times 6.6'$ ($0.79''$
pixel$^{-1}$), and the CTIO 1.3~m observations were obtained using
the dual-channel optical/near-infrared (IR) camera ANDICAM having an
optical FOV  of $6.3 \times 6.3'$ ($0.37''$ pixel$^{-1}$) and an IR
FOV of $2.3' \times 2.3'$ ($0.27''$ pixel$^{-1}$). Broad-band $BVRI$
images were taken with all three telescopes, while the TNT also
followed SN 2006X in the $U$ band and the CTIO 1.3~m telescope
sampled the $J$ and $K$ bands.

As shown in Figure 1, SN 2006X is located between a relatively
bright foreground star and the inner edge of one spiral arm of NGC
4321; thus, light contamination needs to be taken into account when
measuring the SN flux. One way to do this is to apply galaxy
subtraction. Template images of NGC 4321 in various bands were taken
with KAIT and TNT on 2007 March 14 and 16 (respectively), roughly
one year after the discovery of SN 2006X. For the CTIO 1.3~m
observations, we used pre-explosion $B$ and $V$ images of NGC 4321
obtained on 24 April 2000 with the Apache Point Observatory 3.5~m
telescope in New Mexico.

To perform image subtraction, the image containing SN 2006X is first
geometrically registered to the corresponding template image. Next,
the total fluxes of corresponding foreground stars in the SN and
template images are compared, and an appropriate scale factor is
applied to the SN image in order to match the template image. Also,
the point-spread functions (PSFs) of the two images are convolved to
match. The template is then subtracted from the SN image, leaving
the SN with a background that is free from host-galaxy
contamination. The final step is to perform standard PSF-fitting (or
aperture) photometry to measure the instrumental magnitudes for the
SN and the local standard stars with the IRAF\footnote{IRAF, the
Image Reduction and Analysis Facility, is distributed by the
National Optical Astronomy Observatory, which is operated by the
Association of Universities for Research in Astronomy, Inc. (AURA)
under cooperative agreement with the National Science Foundation
(NSF).} DAOPHOT package (Stetson 1987). From the late-time $Hubble
Space Telescope$ (HST) Advanced Camera for Survey (ACS) archival
images (Proposal 10991 by Arlin Crotts 2006), we speculate that SN
2006X was $\sim$22.0 mag in $B$ and $\sim$21.5 mag in $V$ about one
year after maximum. This lingering light barely affects the
early-time photometry but it probably results in underestimating the
luminosity of SN 2006X by 3-6\% when the SN becomes as faint as 18
$-$ 19 mag.

The transformation from the instrumental magnitudes to the standard
Johnson $UBV$ (1966) and Kron-Cousins $RI$ (1981) systems is
established by observing, during a photometric night, a series of
Landolt (1992) standards covering a large range of airmasses and
colors. The color terms obtained on different photometric nights
show some differences, and their average values are listed in Table
1 for the filters used in the TNT, KAIT, and CTIO 1.3~m observations
of SN 2006X.

A total of 10 photometric nights (2 for TNT, 3 for KAIT, and 5 for
the CTIO 1.3~m) were used to calibrate 10 local standard stars in
the field of SN 2006X. Table 2 lists the final calibrated $UBVRI$
magnitudes and their uncertainties, while Table 3 presents the
near-IR $J$-band and $K$-band magnitudes. Only one photometric night
was used to do the $U$-band calibration, so the calibrated
magnitudes may have larger errors. The near-IR magnitudes of Star 10
were measured on 14 nights with respect to the Persson et al. (1998)
standards P9144 and LHS2397a. The measured $J$ and $K$ magnitudes
are 0.07 mag brighter than the values from the 2MASS survey, which
obtained $J$ = 14.793(0.025) mag and $K$ = 13.995(0.040) mag.

These local standard stars are then used to transform the
instrumental magnitudes of SN 2006X to the standard $UBVRIJK$
system, and the final results of the photometry are listed in Table
3 and Table 4. Note that the $BVJK$-band data shown for CTIO 1.3~m
telescope include the S-corrections (see details in \S 3.1) which
are listed in Table 5. The estimated error shown in parentheses is a
quadrature sum of uncertainties in the photometry and the
calibrations. The main source of error comes from the photometry,
caused by photon noise and uncertainties in the image subtraction.

\subsection{Spectroscopy}

Spectra of SN 2006X were primarily obtained with the 2.16 m
telescope at BAO using the Cassegrain spectrograph and BAO Faint
Object Spectrograph and Camera (BFOSC), and with the 3.0 m Shane
telescope at Lick Observatory using the Kast double spectrograph
(Miller \& Stone 1993). Two very late-time spectra were also
obtained at the W. M. Keck Observatory: one with the Low Resolution
Imaging Spectrometer (LRIS; Oke et al. 1995) mounted on the 10~m
Keck I telescope, and the other with the Deep Extragalactic Imaging
Multi-Object Spectrograph (DEIMOS) mounted on the 10~m Keck II
telescope. A journal of spectroscopic observations is given in Table
6.

All spectra were reduced using standard IRAF routines (e.g.,Foley
et~al. 2003). Extraction of the SN spectra was carefully performed
to avoid contamination from the nearby star and the spiral arms of
the galaxy. For the Kast observations, flatfields for the red-side
spectra were taken at the position of the object to reduce near-IR
fringing effects. Flux calibration of the spectra was performed by
means of spectrophotometric standard stars observed at similar
airmass on the same night as the SN. The spectra were corrected for
continuum atmospheric extinction using mean extinction curves for
BAO and Lick observatory; also telluric lines were removed from the
Lick (but not BAO) data. For spectra observed at Lick and Keck, the
slit was always aligned along the parallactic angle to avoid
chromatic effects in the data (Filippenko 1982). This procedure was
not adopted at BAO, however, and it may not significantly affect the
continuum shape as the airmass was usually low ($\sim$1.1) when
getting these spectra for SN 2006X. In general, the relative
spectrophotometry is quite good, consistent with the broadband
photometry within about 5\%.

\section{Light Curves and Spectra}

\subsection{Optical and Near-IR Light Curves}

The $UBVRIJK$ light curves of SN 2006X are presented in Figure 2. No
K-corrections have been applied to any of the magnitudes due to the
small redshift of SN 2006X ($z = 0.0053$). For all the bands we have
excellent temporal coverage, especially around maximum light. In
general the optical measurements obtained with TNT, KAIT, and the
CTIO 1.3~m telescope are consistent with each other within $\pm$0.05
mag, except in the $I$ band where differences are more noticeable.
We find that the TNT $I$-band magnitudes are systematically brighter
than those of KAIT by $0.08 \pm 0.02$ mag and also brighter than
those of the CTIO 1.3~m telescope, especially around the dip
immediately after maximum where the discrepancy reaches $\sim$0.2
mag. This larger difference is likely caused by the different
instrument response function in the $I$ band for the different
systems.

Besides the usual color-term corrections, additional magnitude
corrections between instrumental and standard bandpasses
(S-corrections; Stritzinger et al. 2002) are sometimes required for
the filters that are different from those defined by Bessell (1990).
The filters amounted on TNT are standard Johnson-Cousins $UBVRI$
system, however, the filter transmission curves are unavailable. We
obtained the S-corrections for observations of the CTIO 1.3~m
telescope (see Table 5), except in $R$ and $I$ bands where the
corrections were found to actually increase the systematic
differences of the photometry obtained with different telescopes
(e.g., Kriscusnias et al. 2003). We thus have chosen to let the
ANDICAM $R$ and $I$ photometry stay as they are. The S-corrections
for the filters closer to Bessell's recipe are generally small in
$BVR$ (e.g., $\lesssim$ 0.05 mag) but probably significantly larger
in $U$ and $I$ (e.g., Pastorello et al. 2007). This is demonstrated
by the consistency of the raw magnitudes of TNT and KAIT with the
S-corrected magnitudes of the CTIO 1.3~m in $B$ and $V$ bands within
their errors (see Fig.2). To account for the $I$-band magnitude
difference between the TNT, KAIT, and the CTIO 1~m telescope, we
assign a systematic error of 0.10 mag for all the $I$-band
magnitudes in the following analysis.

A polynomial fit to the $B-$band light curve around maximum yields
$B_{max} = 15.40 \pm 0.03$ mag on JD 2,453,786.17 $\pm$ 0.35 (2006
February 19.93). This means that our observations started from
$-$11.27~d and extended to +116.32~d with respect to the time of $B$
maximum. Likewise, the $V$ light curve reached a peak magnitude of
$14.04 \pm 0.02$ on JD 2,453,789.11 $\pm$ 0.29, about 2.9~d after
$B$ maximum. The fitting parameters for the maxima in the other
bands are presented in Table 7, together with the late time decline
rates (see below). From the $B$ and $V$ light curves we derived an
observed $\Delta m_{15} (B) = 1.17 \pm 0.05$ mag\footnote{The true
$\Delta m_{15}(B)$ for SN 2006X is $1.31 \pm 0.05$, taking into
account the reddening effect on the light-curve shape (Phillips et
al. 1999).} and $B_{max}-V_{max} = 1.36 \pm 0.04$ mag. We also
measured the $B - V$ color at 12~d after $B$ maximum to be $1.83 \pm
0.05$ mag. These colors are much redder than the intrinsic values
shown by other SNe~Ia with similar $\Delta m_{15} (B)$ (see more
discussion in \S 3.3), suggesting that SN 2006X suffered significant
line-of-sight reddening.

In Figures 3$-$8 we compare the $UBVRIJK$ light curves of SN 2006X
with those of other well-observed, nearby SNe~Ia having similar
$\Delta m_{15}$ values, including SNe 1994D ($\Delta m_{15} = 1.27$;
Richmond et al. 1995; Patat et al. 1996), 1998bu ($\Delta m_{15}$ =
1.04; Jha et al. 1999), 2002bo ($\Delta m_{15}$ = 1.15; Benetti et
al. 2004; Krisciunas et al. 2004), 2002er ($\Delta m_{15}$ = 1.32;
Pignata et al. 2004), 2003cg ($\Delta m_{15}$ = 1.25; ER06), and
2005cf ($\Delta m_{15}$ = 1.12; Pastorello et al. 2007). The light
curves of SN 1984A ($\Delta m_{15}$ = 1.20; Barbon et al. 1989),
though not as well observed as those of the other SNe in the
comparison, are also included because SN 1984A shares many similar
properties with SN 2006X, especially in the spectroscopy and
late-time photometry. All the light curves have been normalized to
the epoch of their $B$-band maximum and shifted in peak magnitude to
match those of SN 2006X.

Figure 3 shows that the observed $U$-band light curve of SN 2006X is
slightly broader than those of the other SNe~Ia around maximum
light, but is comparable to that of SN 2003cg which also suffered
significant reddening. Two effects, both associated with large
amounts of reddening, might explain why SNe 2003cg and 2006X are
broader than other SNe in comparison. As demonstrated by Phillips et
al. (1999) and Jha et al. (2007), the total-to-selective absorption
ratio, $R_{\lambda}$, may vary with the supernova phase. For
instance the true value of $R_{U}$ at 15~d past maximum is found to
be smaller than that near maximum by about 0.15 (Jha et al. 2007).
This would subsequently broaden the light curves of the reddened SNe
by $\sim0.15 E(B-V)$. On the other hand, the change of the overall
spectral shape due to the high reddening may shift the effective
bandpass at the shorter wavelengths to the longer wavelengths that
are characteristic of broader light-curve peaks (see ER06 for a
similar argument for SN 2003cg). This may also contribute to the
broader peak in $U$. Another possible feature of SN 2006X in the $U$
band is the late-time light curve, which despite the large error bar
for each measurement, seems to be brighter and possibly declines
more slowly than the other SNe.

Figure 4 shows the comparison in the $B$ band. While the light
curves near maximum are similar to each other, they diverge at late
times. Compared to other SNe~Ia, SNe 2006X, 1984A, and 2002bo have
relatively higher (by 0.3$-$0.5 mag) luminosity after $t = 40$~d,
and their light curves decline more slowly. We measured a late-time
decline rate of $\beta = 0.92 \pm 0.05$ mag (100~d)$^{-1}$ for SN
2006X at $t = 40 - 117$~d, which is comparable to the radioactive
decay rate of 0.98 mag (100~d)$^{-1}$ for the Co$\rightarrow$Fe
decay. The decline rates for SNe 1984A and 2002bo are slightly
higher, $\beta = 1.14 \pm 0.06$ mag (100~d)$^{-1}$ and $1.17 \pm
0.10$ mag (100~d)$^{-1}$, respectively.  All these decline rates are
significantly smaller than the typical value of a normal SN~Ia,
$\beta = 1.40 \pm 0.10$ mag (100~d)$^{-1}$, as exhibited by the
other SNe~Ia in the comparison.

The $V$-band light curve comparison is shown in Figure 5. At around
maximum, SN 2006X is slightly broader, and shows a more prominent
"shoulder" feature at $t \approx 3$ weeks than other SNe~Ia. Part of
the reason for this may be that compared to other SNe~Ia, SN 2006X
has higher reddening, and the effective $V$ bandpass is shifted more
toward the $R$ band, which is broader than $V$ and has a prominent
shoulder feature. It is also possible that there is an intrinsic
scatter in the strength of the shoulder feature in the $V$-band
light curves of SNe~Ia. In the same manner as the definition of
$\Delta m_{15} (B)$, we computed $\Delta m_{15}(V) = 0.63 \pm 0.03$
mag for SN 2006X. This is similar to that of SNe 2003cg, 1998aq, and
2005cf [$\Delta m_{15}(V) = 0.63 \pm 0.05$ mag, $0.67 \pm 0.05$, and
$0.69 \pm 0.03$ mag, respectively], but smaller than that of SNe
1994D and 2002er [$\Delta m_{15}(V) = 0.85 \pm 0.06$ mag and $0.78
\pm 0.04$ mag, respectively]. At late times, SN 2006X has a similar
evolution as SN 2002bo, and both are brighter than other comparison
SNe~Ia. There are no obvious differences in the late-time decline
rates among the SNe.

As shown in Figures 6 and 7, SN 2006X exhibits a more prominent
"plateau" feature in the $R$ band and a stronger secondary maximum
in the $I$ band than other SNe~Ia. The $R$ and $I$ light curves of
SNe~Ia are similar to each other before $t = 10$~d, but diverge
after that. There are considerable differences in $D_{plateau}$, the
duration of the plateau phase in the $R$ band\footnote{D$_{plateau}$
measures the interval of the two inflection times, $t_{1}$ and
$t_{2}$, of the post-maximum $R$-band light curves. The inflection
time is defined as the moment when the first derivative at the
transition phase S = $|dm/dt|$ is minimal (e.g., Elmhamdi et al.
2003).} and the time when the $I$ band reaches the second maximum
[$t(I_{max})$]: SN 2006X has $D_{plateau} \approx 13$~d, and
$t(I_{max}) = 28$~d, while SN 1994D has $D_{plateau} \approx 9$~d
and $t(I_{max}) = 20$~d. At late times, SN 2006X has an evolution
similar to that of SN 2002bo in the $R$ band, and they are brighter
than other SNe~Ia. The two SNe, however, have different decline
rates in the $I$ band, and SN 2006X is the brightest object in the
comparison.

In Figure 8, the $J$-band and $K$-band light curves of SN 2006X are
compared with those of SNe 1998bu, 2002bo, 2003cg, and 2005cf. The
light curves of the well-observed SN 2001el ($\Delta m_{15} = 1.10
\pm 0.03$ mag; Krisciunas et al. 2003) are also overplotted (see the
dotted curves). As in the optical bands, the overall light curves of
SN 2006X in $J$ and $K$ resemble closely those of SN 2002bo. After
the initial peak, both of these SNe show a stronger secondary
maximum feature in the $JK$ bands than other SNe~Ia in the
comparison. In SN 2001el, the $J$ second-peak feature is somewhat
less pronounced; the light curve remains relatively flat from day
+11 to +31. Note that there is marginal evidence for a third maximum
(or shoulder) in the near-IR light curves of SN 2006X at $\sim$70~d
after $B$ maximum. Kasen (2006) has come up with an explanation for
the secondary maximum and has predicted that some SNe~Ia might show
a third maximum at $\sim$80~d. The variation of the secondary
maximum of the near-IR light curves can be related to the abundance
stratification in SNe~Ia, the concentration of iron-group elements,
and the progenitor metallicity (Kasen 2006).

\subsection{Optical and Near-IR Color Curves}

The optical color curves of SN 2006X ($U-B$, $B-V$, $V-R$, and
$V-I$) are presented in Figure 9. The colors of SN 2006X are much
redder than those of the normal Ia SN 2005cf (the triangles) with
negligible reddening, suggesting a significant amount of reddening
toward SN 2006X. Also overplotted in these panels are the color
curves of SNe 1994D, 1998bu, 2002bo, 2002er, 2003cg, all arbitrarily
shifted to match the observed colors of SN 2006X at $B$ maximum.

The $U - B$ color (Fig. 9a) of SN 2006X does not become
progressively redder in a linear fashion after $B$ maximum as the
other SNe~Ia do, but has a shoulder or plateau phase for about two
weeks. A similar feature is observed in the heavily extinguished SN
2003cg. It is unclear whether this feature is caused by the high
reddening of these two objects. After $t \approx 4$ weeks, the $U -
B$ colors of SN 2006X seem to be bluer than those of the comparison
SNe~Ia, but the error bars of the measurements are relatively large
due to the poor $U$-band data quality at late times.

The $B - V$ colors of all SNe in the plot (Fig. 9b) have a similar
evolution before $t \approx 29$~d, although SN 2006X reached the
bluest color a bit earlier, e.g. at $\sim-$8.0 days rather than
$\sim-$5 days for the other comparison SNe. All the SNe~Ia reach the
reddest color at $t \approx 29$~d, although there are some
differences in the time to reach the reddest color. The peak $B - V$
colors of the SNe~Ia with smaller decline rates (or $\Delta m_{15}$)
might occur later than those of the faster decliners (see also
Figure 1 in Wang et al. 2005). We note that the $B - V$ color of SN
2003cg is redder than that of other SNe~Ia by about 0.2 mag at the
reddest peak according to the data from ER06. However, the
photometric data based on the unpublished KAIT data of SN 2003cg do
not have such a discrepancy. Compared with other SNe~Ia, the most
notable difference in the $B - V$ color evolution of SN 2006X is
after $t \approx 29$~d: SN 2006X became bluer in a much faster pace,
and showed an apparent difference from the Lira-Phillips relation
(the dash-dotted line in the figure; see Lira 1995 and Phillips et
al. 1999). Note also that the $B - V$ color of SN 1994D, although
not well observed during this phase, suggests a faster evolution
than the Lira-Phillips relation. Thus, as discussed in the more
extreme case of the peculiar SN Ia 2000cx (Li et al. 2001), although
the Lira-Phillips relation provides a method to estimate the
extinction toward SNe~Ia and may work in a statistical dataset, it
should be used with caution for any individual SN~Ia.

Given the relatively large differences seen in the $U - B$ and $B -
V$ color evolution between SN 2006X and other SNe~Ia in the
comparison, it is surprising that all SNe~Ia have similar $V - R$
color evolution except at very early times (Fig. 9c). The $V - R$
color of different SNe~Ia reaches the bluest point at $\sim$12~d
after $B$ maximum, then become progressively redder and reaches the
maximum around day +31. After the peak the $V - R$ color curve moves
gradually toward bluer colors. The similar evolution in the $V - R$
color is related to the closer match of the $V$ and $R$ photometry
and also a smaller reddening effect.

The $V - I$ color curve of SN 2006X (Fig. 9d) exhibits a behavior
which is very similar to those of the comparison SNe. All these SNe
reached a minimum color at about 9~d after $B$ maximum. The
following red peak occurred at around day +31. The only exception is
SN 1994D, which showed a rapid increase in $V - I$ after the minima
and then reached the peak value 4~d earlier than other SNe~Ia.

Figure 10 shows the observed $V - J$ and $V - K$ colors of SN 2006X,
along with those of SNe 1998bu, 2001el, 2002bo, 2003cg, and 2005cf
(without color shift). All SNe~Ia exhibit a deep minimum at $t
\approx 8$--11~d, followed by a maximum at around day +32.
Nevertheless the $V$ minus near-$IR$ color of the comparison SNe~Ia
shows noticeable scatter, especially in $V - J$ where the magnitude
contrast between the valley and peak can differ by more than 0.6
mag. The dashed curves in Fig. 10 represent the average loci of the
$V-J$ and $V-K$ colors for normal SNe Ia (Krisciunas et al. 2000)
that were reddened (shifted upward in Fig.10) by 1.72 mag and 1.95
mag, respectively. With respect to the average color curve, the $V -
J$ and $V - K$ colors of SN 2002bo seem to better match those of SN
2006X.

In general, the overall color evolution of SN 2006X resembles most
closely those of SN 2002bo and SN 2002er. Other comparison SNe~Ia
may match well with SN 2006X in $V - R$ and $V - I$, but they show
larger differences than SN 2006X in $B - V$, $V - NIR$. Both SN
2006X and SN 2003cg exhibited an atypical ``shoulder" feature in $U
- B$ for about two weeks immediately after $B$ maximum.

\subsection{The Reddening Toward SN 2006X}

The Galactic extinction toward NGC 4321/SN 2006X is
$A^{\rm{gal}}_{V}=0.087$ mag (Schlegel et al. 1998), corresponding
to a color excess of $E(B - V)$ = 0.026 mag [adopting the standard
reddening law of Cardelli, Clayton, \& Mathis (1989)]. As suggested
by the observed color curves of SN 2006X shown in Figures 9 and 10,
SN 2006X may suffer a large amount of extinction within its host
galaxy. In this section, we use two methods, global color-curve
fitting and the $\Delta m_{15}$ method, to derive the host-galaxy
reddening for SN 2006X. Both methods assume SN 2006X has colors
similar to those of the other SNe~Ia with either similar evolution
in some colors, or comparable $\Delta m_{15}$ values. Given the
somewhat unusual evolution in the $U - B$ (a plateau phase) and $B -
V$ (an apparent deviation from the Lira-Phillips relation) colors,
however, we caution that SN 2006X may have peculiar intrinsic
colors, in which case it would be difficult to estimate the
host-galaxy reddening toward SN 2006X.

From the comparison of color curves discussed in \S 3.2, we find
that SN 2006X has a color evolution similar to that of SN 2002bo in
$B-V$ and $V - R,I,J,K$, and to that of SN 2002er in $B - V$ and $V
- R,I$. The host-galaxy reddenings of SNe 2002bo and 2002er have
been determined by different groups (Reindl et al. 2005; Wang et al.
2006; Jha et al. 2007) to be $E(B - V)_{host} = 0.40 \pm 0.04$ mag
and $0.16 \pm 0.03$ mag, respectively. We remove both the Galactic
and the host-galaxy reddening from SNe 2002bo and 2002er and create
their intrinsic color curves, and then apply different amounts of
color excesses to the colors of SN 2006X and find the best global
fit to these intrinsic color curves. We limit the fit to data with
$t \lesssim +40$~d since the $B - V$ color of SN 2006X shows an
apparently distinct evolution at late times. This procedure yields
the average color excesses $E(B - V) = 1.41 \pm 0.07$ mag, $E(V - R)
= 0.61 \pm 0.05$ mag, $E(V - I) = 1.26 \pm 0.05$ mag, $E(V - J) =
1.76 \pm 0.09$ mag, and $E(V - K) = 1.95 \pm 0.09$ mag.

Phillips et al. (1999) and Wang et al. (2005, 2006) proposed
correlations between the light-curve width parameter $\Delta m_{15}$
and the intrinsic  $B_{max} - V_{max}$ value (or $C_{max}$) or the
$B - V$ color at 12~d after $B$ maximum [$(B - V)_{12}$;  or $\Delta
C_{12}$], which can extend to other optical colors and even to the
$V$ minus near-$IR$ color (e.g., see Fig. 13 in Krisciunas et al.
2004). We use these two correlations to estimate the host-galaxy
reddening of SN 2006X. Phillips et al. (1999) found that large
reddening changes the observed $\Delta m_{15}$, with $\Delta
m_{15}(B)_{true} \approx \Delta m_{15}(B)_{obs} + 0.1 E(B -
V)_{true}$, where $\Delta m_{15}(B)_{true}$ and $\Delta
m_{15}(B)_{obs}$ are the intrinsic and observed $\Delta m_{15}(B)$
values, respectively, and $E(B - V)_{true}$ is the total reddening
toward the SN. Elias-Rosa et al. (2006) suggest that this relation
works well for the high-reddening case of SN 2003cg. As we use
$\Delta m_{15}(B)_{true}$ to infer the intrinsic $C_{max}$ and
$\Delta C_{12}$, and derive the reddening toward SN 2006X, this is a
loop process; we set $E(B - V)$ as a free parameter, and iterate the
fitting process until a convergence is achieved.

The color excesses derived from these two methods are consistent
with each other within the uncertainties, and they are summarized in
Table 8. The weighted average value of $E(B - V)$ is $1.42 \pm 0.04$
mag, indicating that SN 2006X suffers heavy host-galaxy reddening.

Another independent way to estimate the reddening toward SN 2006X is
to measure the equivalent width (EW) of the interstellar Na~I~D
doublet, which is proposed to correlate with the amount of
line-of-sight dust (Barbon 1990; Munari \& Zwitter 1997). Due to the
line blending and the low resolution, we only use the spectra at
days +13, +68, and +75 to estimate the EW of Na~I~D. This yields an
average EW $\approx$ 1.7 \AA \  for SN 2006X (but see Patat et al.
2007 for measurements of high-resolution spectra and the possible
variation of the Na~I~D EW). Using the prescription of Barbon
(1990), $E(B - V) = 0.25 \times$ EW (Na~I~D), we derive a reddening
of $E(B-V) \approx 0.43$ mag toward SN 2006X. However, the
prescription by Munari \& Zwitter (1997), with $E(B - V) = 0.36
\times$ EW (Na~I~D), suggests a larger reddening of $E(B - V)
\approx 0.61$ mag. Both determinations are significantly smaller
than that derived from the photometric method. Due to the large
uncertainty in the correlation between the EW of the Na~I~D
interstellar absorption and the reddening, we adopt the reddening
derived from the photometric method in the following analysis.

\subsection{The Extinction Coefficient $R_{V}$}

Assuming that the dust surrounding SN 2006X conforms with the
extinction law proposed by Cardelli et al. (1989), we can set
constraints on the value of $R_{V}$. Cardelli et al. provide an
analytic expression for the average extinction law,
$A_{\lambda}/A_{V} = a_{\lambda} + b_{\lambda}/R_{V}$ (where
$a_{\lambda}$ and $b_{\lambda}$ are wavelength-dependent
parameters), which well reproduce the average extinction curve of
the Galaxy with $R_{V} = 3.1$. The value of $R_{V}$ for SN 2006X can
be obtained by simultaneously matching different color excesses
derived in \S 3.3 using our adopted $E(B - V)$ of $1.42 \pm 0.04$
mag.

Figure 11 shows that the best fit to the mean color excesses
($UBVRIJK$ minus $V$) is achieved with $R_{V} = 1.48 \pm 0.06$,
which is much smaller than the standard value of 3.1. This gives a
host-galaxy visual extinction $A_{V}$(host) = 2.10 mag for SN 2006X.
Excluding the color excesses derived from the near-IR bands in the
fit yields the optimal $R_{V}$ = 1.58$\pm$0.15 and $A_{V}$(host) =
2.24 mag. Allowing the best fit $E(B - V)_{host}$ to vary marginally
alters these results. Fitting the color excesses of SN 2006X using
the standard extinction law ($R_{V} = 3.1$) yields a reduced
$\chi^{2}$ of 284.9 and 21.4 for the $UBVRIJK$ and the $UBVRI$
colors, respectively. This indicates that the dust surrounding SN
2006X is quite different from that observed in the Galaxy, probably
of a much smaller grain size than that of typical interstellar dust.
Note that the derived $R_{V}$ value is just the average value; the
true value may actually show small variance with supernova phase.
For comparison, we also determined the host-galaxy reddening of SN
2006X using the MLCS2k2 method (Jha et al. 2007), which gives
$A_{V}$(host) = 2.21 mag and $R_{V} = 1.58$, quite consistent with
our determinations with the optical colors.

Besides SN 2006X, we note that other heavily extinguished SNe~Ia
also tend to suffer reddening by abnormal dust. In Table 9 we list
the color excesses and $R_{V}$ derived for several SNe~Ia with $E(B
- V)_{host} \gtrsim 0.50$ mag published in the literature. While the
$R_{V}$ values of SNe 1995E and 2000ce are within $2\sigma$ of 3.1
given their relatively large error bars, the $R_{V}$ values of the
other SNe are considerably smaller than 3.1. It is interesting to
investigate whether the low $R_{V}$ derived for these highly
extinguished SNe~Ia is universal for all SNe~Ia or an exception for
a few events. Wang et al. (2006) derived an average $R_{V} = 2.32
\pm 0.12$ for Hubble-flow SNe~Ia. In the MLCS2k2 method, an $R_{V}$
value lower than 3.1 is often required to fit heavily reddened
SNe~Ia. More SNe~Ia with well-observed multicolor light curves will
help diagnose the properties of the dust surrounding SNe~Ia. We note
that although the infrared light curves themselves suffer little
from the dust absorption, their color differences depend more
sensitively on the ratio $R_{\lambda}$ than the optical colors (as
indicated by the difference between the two curves with different
$R_{V}$ in Fig. 11), so they are particularly suitable for studying
the $R_{V}$ of SNe~Ia.

\section{Optical Spectra}

Twelve optical spectra of SN 2006X obtained at BAO and Lick
Observatory, spanning $t = -0.9$ to +97.8~d, are displayed in Figure
12. Two late-epoch Keck spectra, taken at days +277.0 and +307.0,
are also shown in the plot. The spectral evolution generally follows
that of a typical SN~Ia near maximum brightness, but with more
pronounced absorption features at 3700~\AA\ due to Ca~II H\&K,
6020~\AA\ due to Si~II $\lambda$6355, and 8100~\AA\ due to the Ca~II
near-IR triplet. Other features at early times include the S~II
lines near 5400~\AA\ and the blended lines of Fe~II and Si~II near
4500~\AA. Unlike a normal SN~Ia, however, the overall continuum of
SN 2006X appears quite flat within the first 2 weeks after $B$
maximum because of its high reddening. We discuss the detailed
spectral evolution of SN 2006X in the following sections.

\subsection{Spectroscopic Evolution at Early Times}

In Figures 13--15, we compare the spectra of SN 2006X with those of
SNe 1984A (Barbon et al. 1989), 1994D (Filippenko 1997), 2002bo
(Benetti et al. 2004), 2002er, and 2003cg (ER06) at several epochs
($t \approx 0$~d, one week, and one month past $B$ maximum). The
spectra of SN 2002er are taken from our unpublished spectral
library. All of the spectra have been corrected for reddening and
redshift. For SN 2006X, $E(B - V)_{host} = 1.42$ mag and $R_{V} =
1.48$ are used, while $E(B - V)_{host} = 1.33$ mag and $R_{V} = 1.8$
are used for SN 2003cg (ER06). For the other SNe, the host galaxy
reddening is taken from the literature (e.g., Reindl et~al 2005,
Wang et~al. 2006, and Jha et~al. 2007) and extinctions are corrected
using the standard extinction law with $R_{V} = 3.1$. The line
identifications adopted here are taken from Branch et al. (2005,
2006).

At $t \approx 0$~d, the spectra of SN 2006X are characterized by
lines of the singly ionized intermediate-mass elements (IMEs):
Ca~II, Si II, and Mg II (Fig. 13). The width and strength of these
lines are comparable to those seen in SNe 1984A and 2002bo, but
significantly stronger than those of SNe 1994D, 2002er, and 2003cg.
The absorption minima due to Si~II $\lambda$6355 is highly
blueshifted to $\sim$ 6020~\AA\ around maximum light, suggesting a
high photospheric expansion velocity for SN 2006X. The strong Ca~II
H\&K feature, blended with Si~II $\lambda$3858, is more similar to
that of SNe 2002bo and 2002er than to the double-dipped feature seen
in SNe 1994D and 2003cg. The presence or absence of such a
double-dipped feature is probably related to the relative strength
of Ca~II and Si~II as suggested by Lentz et al. (2000). In the
4000$-$4500~\AA\ wavelength range, SNe 2006X, 1984A, and 2002bo
share a similar feature with a strong absorption at around 4300~\AA,
probably owing to a blend of Mg~II $\lambda$4481 and Fe~III
$\lambda$4404. The weak feature at $\sim$4400~\AA\ as seen in SNe
1994D, 2002er, and 2003cg is probably due to the Si~III
$\lambda\lambda$4553, 4568 blend, which is more sensitive to the
temperature in the photosphere. Its absence in the spectra of SNe
2006X, 2002bo, and 1984A suggests a cool temperature. The double
S~II lines at 5400~\AA\ and 5600~\AA\ (the ``W''-shaped feature) in
SN 2006X are similar to those in the comparison SNe, but are less
pronounced. The Si~II $\lambda$5972 absorption feature in SN 2006X
is very weak, and its strength is difficult to measure.
Nevertheless, we measured the potential luminosity indicator
parameter R(Si~II) (defined as the line strength ratio of Si~II
$\lambda$5972 to Si II $\lambda$6355; Nugent et al. 1995) as $0.12
\pm 0.06$ for SN 2006X, which is much lower than in the other SNe~Ia
having similar values of $\Delta m_{15}$ (Fig.3 of Benetti et al
2005).

At $t \approx 1$ week (Fig. 14), the spectra of all the SNe~Ia are
still dominated by the Si~II, Ca~II, and S~II lines with increasing
contribution from the iron-group elements. The W-shaped S~II lines
have almost vanished in SNe 2006X and 1984A, and appears as an
asymmetric absorption trough in SN 2002bo. The absorption at around
4400~\AA\ is probably due to a blend of Fe~III $\lambda$4404, Fe~II
$\lambda$4555, and Mg~II $\lambda$4481, and appears at a higher
expansion velocity (shorter wavelength) in SNe 2002bo, 1984A, and
2006X than in the other SNe~Ia. The Fe~II and Si~II features in the
range 4700--5000~\AA\ show less substructure in SNe 2006X, 2002bo,
and 1984A, while their Si~II $\lambda$5051 line becomes apparently
weaker than that of the other comparison SNe~Ia. This probably
relates to the high expansion velocity which may smear out the
weaker features, or it suggests faster evolution of the S and Si in
the high-velocity SNe Ia. By $t \approx 1$ week the O~I
$\lambda$7773 line strengthens in other SNe~Ia but is still rather
weak in SN 2006X.

At $t \approx 1$ month (Fig. 15), numerous Fe~II lines appear in the
spectra of SNe~Ia. The Na~I~D and Si~II $\lambda$5972 blend is
progressively stronger. The Si~II $\lambda$6355 trough becomes
severely affected by the Fe~II $\lambda\lambda$6238, 6248 and Fe~II
$\lambda\lambda$6456, 6518 lines, and is barely visible in SN 2006X
and SN 1984A, but is still present in the spectra of the other
comparison SNe (see the small dip around 6150\AA\ in Fig.15 ). The
Ca~II near-IR triplet feature is stronger, and appears at a higher
expansion velocity in SNe 2002bo and 2006X than in the other SNe~Ia.
The O~I line is comparable to that seen in other SNe~Ia. It is
impressive that the overall appearance and evolution of the spectra
of SN 2006X are extremely similar to those of SN 1984A and SN 2002bo
at all three epochs (although the spectra of SN 1984A have a limited
spectral range). Nevertheless, it is worth pointing out that by $t
\approx 1$ month the spectral flux at $\lambda$ $<$ 4000~\AA\
appears relatively higher than that of the comparison SNe~Ia,
although the spectrum at this region had a poorer signal-to-noise
ratio. The higher spectral flux below $\lambda$ $\sim$ 4000\AA\
seems to be also consistent with the bluer $U - B$ color inferred
from the photometry at similar phase. The spectrum near maximum
light does not show such a near-UV energy excess. It is unclear when
this variation started to occur in SN 2006X, as our spectra taken
during the first month past maximum light did not cover the near-UV
wavelengths.

\subsection{Expansion Velocity of the Ejecta}
As the photosphere recedes into deeper and more slowly moving
ejecta, the P-Cygni absorption minima of the spectral features in a
SN~Ia spectrum gradually shifts redward with time (see also Fig.12).
Thus the location of the blueshifted minima can in principle trace
the photospheric velocity. Indeed, the photospheric velocity
($v_{exp}$) is best measured by the weakest lines, e.g. S~II
$\lambda$ 5640 (Jeffery \& Branch 1990, Mazzali et al. 1993, Patat
et al. 1996), although it is usually difficult to measure these weak
lines due to blending. The velocities inferred from the strong lines
(e.g., Si~II $\lambda$6355) represent approximately the $v_{exp}$ at
early phase when the amount of the material above the photosphere is
small, but they may not properly indicate of the real $v_{exp}$ as
the photosphere falls in the regions where the strong lines could
form over a large velocity range due to a flatter density gradient.
On account of rather weak S~II lines after maximum in SN 2006X, we
only examined the velocity evolution of Si~II $\lambda$6355 and the
Fe~II/Fe~III lines.

The derived $v_{exp}$ values from Si~II $\lambda$6355 of SN 2006X as
a function of time are shown in Figure 16, together with those of
the comparison SNe~Ia. The $v_{exp}$ reported for an early-epoch
spectrum, taken on 8.35 February (corresponding to $t \approx
-11.3$~d) by Quimby, Brown, \& Gerardy (2006), is also included in
the plot (shown as the open star). All velocities have been
corrected for the redshifts of their respective host galaxies. As
can be seen, SN 2006X shows the highest expansion velocities among
all SNe in the comparison, with $v_{exp}$ = 20,700 km s$^{-1}$ at
$t= -11.3$~d and $v_{exp}$ = 15,700 km s$^{-1}$ at $t= -1.2$~d,
while the typical value for most SNe~Ia is around 11,000 km s$^{-1}$
at maximum brightness (Filippenko 1997; see also Figure 1 in Benetti
et al. 2005). The $v_{exp}$ measured from Ca~II H\&K and S~II
$\lambda$5640 near maximum are about 20,500 km s$^{-1}$ and 11,800
km s$^{-1}$, respectively, which are also significantly higher than
those measured for the other SNe~Ia. SN 1984A and SN 2002bo showed
similar, but less pronounced, high-velocity features of the IMEs
(Si~II, S~II, and Ca~II). Following Benetti et al. (2005), we
calculate the velocity gradient $\dot{v}$ of Si~II $\lambda$6355 for
SN 2006X during the period from $t \approx 0$ to $t \approx 30$~d as
$123 \pm 10$ km s$^{-1}$ d$^{-1}$, which puts SN 2006X in the group
of normal SNe~Ia with high velocity gradients (HVGs). Other SNe~Ia
in the HVG group include SNe 1984A and 2002bo discussed in this
paper, SNe 1983G, 1989A, 1997bp, and 2002dj (Benetti et al. 2005),
and 2004dt (Wang L et al. 2006b).

The Fe~II and Fe~III lines are sometimes used to measure the
expansion velocities of the inner ejecta of Fe, providing additional
clues to the nature of SN~Ia explosions (e.g., Li et al. 2001). The
values of $v_{exp}$ determined from Fe~II $\lambda$4555 and/or
Fe~III $\lambda$4404 lines of different SNe~Ia are plotted in Figure
17. At early phases the high-excitation Fe~III $\lambda$4404 line is
possibly contaminated by Mg~II $\lambda$4481 and hence may not give
a reliable measurement of the $v_{exp}$. By $t \approx$ 2--3 weeks,
the Fe~II $\lambda$4555 line becomes stronger and gradually
dominates the absorption feature near 4300--4400~\AA. This
absorption feature yields $v_{exp} \approx$ 10,000 km s$^{-1}$ for
SN 2006X at $t \approx 1$ month, which is about 3000 km s$^{-1}$
higher than that for SNe 1989B and 1994D and comparable to that for
SN 2000cx (Li et al. 2001). Likewise, the Fe~II lines of SNe 1984A
and 2002bo display high-velocity behavior. As with Si~II
$\lambda$6355, the Fe~II feature in these high-velocity events shows
more rapid evolution, with larger velocity gradients than in other
comparison SNe~Ia. The values of $v_{exp}$ become much closer to
each other when SNe~Ia start entering the nebular phase.

The origin of the high-velocity features seen in these SNe~Ia is
hotly debated. Lentz et al. (2000) find that the strength, profile,
and velocity of Si~II $\lambda$6355 is a function of metallicity.
The blueward shift of the Si~II feature is found to increase with
higher metallicity. However, Benetti et al. (2004) found that
increasing the normal metallicity in the C+O layer by a factor of 10
in the canonical deflagration model W7 is not adequate for
explaining the large $v_{exp}$ seen in SN 2002bo. It was also
proposed that the high $v_{exp}$ in these SNe~Ia may result from
delayed-detonation explosions with the transition density from a
deflagration to a detonation as the controlling parameter for the
internal dispersion of $v_{exp}$ (Lentz et al. 2001; Benetti et al.
2004). With detailed non-LTE calculations, Lentz et al. (2001) find
that some delayed-detonation models provide reasonable
approximations to the very high-velocity feature of SN 1984A. The
increase in density of IMEs at higher velocities is thus responsible
for the larger blueshifts of the line minimum in the spectra and
hence the higher measured expansion velocities.

\subsection{Spectra in the Nebular Phase}

Since the $B$-band and probably $V$-band light curves of SN 2006X
declined more slowly at late times than those of the other
comparison SNe~Ia, it is necessary to examine its late-time spectral
behavior in detail. Our collection of nebular-phase spectra of SN
2006X is presented in Figures 18 and 19, together with the spectra
of some comparison SNe~Ia at similar epochs. Proper corrections for
the reddening and redshift have been applied to all the spectra.

In the early nebular phase, at $t \approx 3$ months, the spectra in
Figure 18 are quite similar to each other but have subtle
differences. The spectra are dominated by various iron lines and
have similar relative strengths. The Ca~II near-IR triplet is still
the strongest feature in the earlier nebular spectra. Although the
late-time spectra of SN 2006X are rather noisy, they do show a
relatively higher flux at $\lambda < 5000$~\AA, and an overall bluer
continuum than the comparison SNe. The most contrast is with SN
2002bo, in which the relative luminosity in $U$ seems to be fainter
than that of SN 2006X by about 1.0 mag. The near-UV excess of SN
2006X might have also occurred in the spectrum at $t \approx 30$~d
since $B$ maximum (see Fig. 15). This is consistent with the bluer
$U - B$ and $B - V$ colors of SN 2006X compared with the other SNe
(Fig. 9).

Two very late-phase nebular spectra, obtained with the Keck 10~m
telescopes at days +277.0 and +307.0, are shown in Fig. 19. The
comparison spectra of SNe 1996X and 1998bu are from the Suspect
online supernova spectrum archive (contributed by Salvo et al. 2001
and Jha et al. 1999; see http://bruford.nhn.ou.edu/$\sim$suspect),
and the spectra of SN 2003du are from Stanishev et al. (2007). The
spectra at this time are dominated by forbidden lines of singly and
doubly ionized Fe and Co. The overall shape of the spectrum of SN
2006X looks similar to that of SN 1998bu, but the profiles and
intensities of some features do show significant differences. The
most pronounced one is the absorption feature at about 6100~\AA,
which is stronger in SN 2006X than in the other SNe~Ia in our
comparison. Other discrepancies include the absorption features near
4200~\AA\ and 4500~\AA, which are prominent in SNe 1996X, 1998bu,
and 2003du but are marginally visible in SN 2006X. We note that both
SN 2006X and SN 1998bu show relatively higher flux below 4500~\AA\
at this phase. The extra flux for SN 1998bu could be caused by a
light echo (Cappellaro et al. 2001; Spyromilio et al. 2004), which
primarily contributes to the spectrum in the blue.

A possible explanation for the abnormal behavior of the
nebular-phase spectra and light curves of SN 2006X in the blue is
the presence of a light echo. As this supernova is found to suffer
significant extinction, it may have occurred in an environment that
is abundant with dust, so the occurrence of a light echo is
potentially expected. Some of the observable effects of a light echo
around a SN Ia include a bluer late-time color, broader spectral
lines, and a brighter tail luminosity (Patat 2006), all of which
agree with the observations of SN 2006X. The unidentified, prominent
feature at $\sim$6100~\AA\ seen in the spectrum at day +307 could be
reminiscent of the earlier-epoch Si~II $\lambda$6355, reflected by
the surrounding dust. Analysis of the light echo in SN 2006X is
given in a separate paper (Wang et al. 2007), in which evidence from
the late-time $HST$ image and spectrum is presented. In particular,
the PSF-subtracted HST image distinctly reveals prominent echo
structure at $\sim 0.05''- 0.13''$ away from the SN. The echo
emission might exist at $<$ 0.05'' from the SN site
(at $\sim$ 2-$\sigma$ level), as suggested by that the echo
inferred from the spectrum is brighter than that resolved in the HST images.
However, the exact location of the possible local echo is unknown due to the
limitation of the image resolution. With the Cepheid distance to M100 $\sim$15.2 Mpc,
the materials illuminated by the resolved LE are found to lie at $\sim$ 27 pc $-$
170 pc from the SN, and it is not clear whether they were
produced from a distinct dust component, e.g. the local
circumstellar dust, or from a continuous dust distribution as with
the outer echo component.

\section{The Distance and Luminosity of SN 2006X}

The host galaxy of SN 2006X, NGC 4321 (M100), is a well-studied
LINER/H~II galaxy (e.g., Ho, Filippenko, \& Sargent 1997). As one of
the largest spiral galaxies in the Virgo cluster, it produced SNe
1901B, 1914A, 1959E, 1979C, and 2006X in roughly the last century.
Ferarrese et al. (1997) reported a Cepheid distance to NGC 4321 of
$m-M = 31.04 \pm 0.17$ mag. A distance of $m-M = 30.91 \pm 0.14$ mag
was published by the {\it HST} Key Project (Freedman et al. 2001),
which we adopt here in our analysis.

\subsection{Absolute Magnitudes and $H_{0}$}

With the Cepheid distance and the reddening derived in the previous
sections, it is straightforward to calculate the absolute magnitudes
of SN 2006X. After correcting for the Galactic reddening of $E(B -
V)_{Gal} = 0.026$ mag with $R_{V} = 3.1$ and the host-galaxy
reddening of $E(B - V)_{host} = 1.42 \pm 0.04$ mag with $R_{V} =
1.48 \pm 0.06$, we derive the $B$-band and $V$-band absolute
magnitudes to be $-19.10 \pm 0.20$ and $-19.06 \pm 0.17$ mag,
respectively. The magnitudes in other bands are listed in Table 10.

To compare with other SNe~Ia, we need to normalize the derived
absolute magnitudes of SN 2006X to a nominal light-curve shape, or
$\Delta m_{15}$ value. Phillips et al. (1993) proposed a relation
between $\Delta m_{15}$ and the peak luminosity of SNe~Ia, and there
are now several different versions available in the literature
(e.g., Hamuy et al. 1996; Phillips et al. 1999; Altavilla et al.
2004). Based on a large sample of SNe~Ia, Prieto et al. (2006)
updated the $M_{max} - \Delta m_{15}$ relation (see their Table 3),
which we adopt to normalize the luminosity of SN 2006X. An
alternative luminosity correction method was proposed by Wang et al.
(2005), who introduced a post-maximum color parameter $\Delta
C_{12}$ which correlates well with the maximum luminosity of SNe~Ia.
The relevant correction coefficients are taken from Table 3 in Wang
et al. (2006).

The normalized luminosities (to $\Delta m_{15} = 1.1$ mag) of SN
2006X from the two methods are consistent with each other, and are
listed in Table 10. The normalized luminosity of SN 2006X is
consistent with those of the fiducial SN~Ia in the $U$, $B$, $V$,
and $R$ bands, while it is fainter by $\sim$0.2--0.3 mag in the $I$,
$J$, and $K$ bands. This may be partly due to the more pronounced
spectral features in the red, such as the broader and deeper
absorption trough of the Ca~II near-IR triplet in SN 2006X than in
other SNe~Ia, as discussed in \S 4.1.

The measured luminosities for SN 2006X from the Cepheid distance to
its host galaxy allow us to determine the Hubble constant via the
formula $H_{0} = 10^{0.2(M+25-\alpha)}$, where $M$ is the absolute
magnitude of SN 2006X and $\alpha$ is the zero point defined by
Hubble-flow SNe~Ia. Using the normalized luminosities calibrated
from the $\Delta C_{12}$ method and the zero point determined by
Wang et al. (2006) from 73 Hubble-flow SNe~Ia, we derive a Hubble
constant (in units of km s$^{-1}$ Mpc$^{-1}$) of 71.1, 73.5, 73.8,
83.9 from the $U$, $B$, $V$, and $I$ bands, respectively. Excluding
the large value obtained with the peculiar $I$-band data, the
average value of $H_{0}$ derived from SN 2006X is $72.8 \pm 8.2$ km
s$^{-1}$ Mpc$^{-1}$ (statistical), which is consistent with the
estimates from other Cepheid-calibrated SNe~Ia (Jha et al. 1999;
Riess et al. 2005; Wang et al. 2006). The statistical error quoted
here consists of the uncertainty in the extinction correction and
the intrinsic luminosity dispersion of SNe~Ia. An analysis of the
uncertainty associated with the Cepheid distances is important but
is beyond the scope of this paper.

\subsection{Bolometric Light Curve and Nickel Mass}

To better understand the overall properties of SN 2006X, we
constructed its quasi-bolometric light curve using our $UBVRIJK$
photometry. For this calculation, we used the normalized passband
transmission curves given by Bessell (1990). The integrated flux in
each band was approximated by the mean flux multiplied by the
effective width of the passband. As we did not get any ultraviolet
(below 3200\AA ) data in SN 2006X, we corrected for the contribution
of $UV$ flux in our calculation using the recipe given by Suntzeff
(1996). Following the procedure in Contardo et al. (2000), we
inferred the corrections for the passbands missing between $U$ and
$I$ from those SNe Ia which have good observations in all the
filters. For the flux beyond $I$ band, we used our own $JK$
data\footnote{The missing $H$-band data may not impact significantly
to the construction of the quasi-bolometric light curve, as the flux
contribution in $H$ inferred from the SNe Ia with good observations
in near $IR$ bands (e.g. SN 2001el, Kricisunas et al. 2003) is
generally found to be smaller than 3\% of the total flux before 80
days after $B$ maximum.} up to $\sim$ 80 days and then applied the
Suntzeff $IR$ correction. The resulting quasi-bolometric light curve
of SN 2006X is shown in Figure 20, together with those of the
comparison SNe~Ia.

The reddening-corrected quasi-bolometric luminosity of SN 2006X is
($1.02 \pm 0.10) \times 10^{43}$ erg s$^{-1}$ at maximum brightness,
comparable to that of most comparison SNe~Ia but slightly fainter
than SN 1998bu. Differences in the peak luminosity are often
considered to be caused by variations in the amount of $^{56}$Ni
synthesized during the explosion. The ``bump'' feature at $t =
20$--32~d is more prominent in SN 2006X than in other comparison
SNe. This is caused by the more pronounced shoulder and ``second
maximum'' seen in the $R$, $I$, and near-IR bands in SN 2006X. We
also note that the contrast between the peak and tail luminosities
of SN 2006X is the smallest among the SNe in the comparison: it has
a peak luminosity in the middle of the group, but the highest tail
luminosity. Efficiency of the trapping of the gamma rays and the
positrons from $^{56}$Co decay (Milne et al. 2001), and small-scale
light echoes or ejecta-CSM interaction, may account for this
difference.

The radioactive $^{56}$Ni synthesized during a SN~Ia explosion is
the primary physical parameter determining the peak luminosity, the
light curve width, and the spectrosocpic/color evolution of the
event (e.g., Kasen et al. 2006). One method to estimate the
synthesized $^{56}$Ni mass is by assuming that the luminosity at
maximum equals the instantaneous energy deposition rate from the
radioactive decay, the so-called ``Arnett law'' (Arnett 1982; Arnett
et al. 1985; Branch 1992). Following Stritzinger \& Leibundgut
(2005), the maximum luminosity produced by the radioactive $^{56}$Ni
can be written as
\begin{equation}
 L_{\rm max} = (6.45e^{\frac{-t_{r}}{8.8}} +
 1.45e^{\frac{-t_{r}}{111.3}})(\frac{M_{\rm Ni}}{M_{\odot}})\times 10^{43}
\,\rm{erg\ s^{-1}}.
\end{equation}
where $t_{r}$ is the rise time of the light curve (e.g., the time
spent by the SN from the explosion to the $B$-band maximum), and
$M_{Ni}$ is the $^{56}$Ni mass (in unit of the solar mass
$M_{\odot}$). Assuming that the luminosity of a SN Ia evolves as an
expanding fireball at very early phase (Goldhaber 1998), the rise
time $t_{r}$ can be derived from the relation (Riess et al. 1999) :
\begin{equation}
L(t) = \varepsilon (t + t_{r})^{2}.
\end{equation}
where $t$ is the elapsed time since maximum and $\varepsilon$ is the
parameter describing the rising rate of the luminosity. Using very
early photometric data in the $R$ band (including the earliest
unfiltered data from IAU Circ. 8667), we derive the rise time to the
$B$ maximum as 18.2$\pm$0.9 d. This is slightly shorter than the
average rise time of 19.5$\pm$0.2 d for a normal SN Ia with $\Delta
m_{15}$ = 1.1 mag (Riess et~al. 1999, but see Strovink 2007 for the
study of a possible dichotomy of the rise time in SNe Ia), but is
consistent with that derived for SN 2002bo ($t_{r}$ = 17.9$\pm$0.5
d; Benetti et~al. 2004) and SN 2005cf ($t_{r}$ = 18.6$\pm$0.4 d;
Pastorello et~al. 2007).

With the peak quasi-bolometric luminosity and the rise time derived
for SN 2006X, we derive a $^{56}$Ni mass of $0.50 \pm
0.05$~M$_{\odot}$. This is within the reasonable range of $^{56}$Ni
masses of normal SNe~Ia. The quoted error bar includes uncertainties
in the rise time and bolometric luminosity.

Table 11 lists all the important parameters derived for SN 2006X in
the previous sections as well as the parameters relevant to its host
galaxy M100.

\section{Discussion and Conclusions}

In this paper we present extensive optical and near-IR photometry,
as well as optical spectroscopy, of the nearby SN Ia 2006X in NGC
4321 (M100). Our observations indicate that SN 2006X is a highly
reddened object with an $R_V$ value much smaller than the canonical
3.1 for average Galactic dust, and has the highest expansion
velocity ever published for a SN~Ia.

Compared with other SNe~Ia included in this paper, SN 2006X has a
broader light curve in the $U$ band, a more prominent ``bump''
feature in the $V$ band, a more pronounced shoulder in the $R$ band,
and a brighter second maximum in the $IJK$ bands.  Relative to the
peak brightness, SN 2006X has a higher luminosity in the $UBVRIJK$
light curves during the nebular phase than other comparison SNe~Ia.
The $B$-band decline rate at late times, $\beta = 0.92 \pm 0.05$ mag
(100~d)$^{-1}$, is much smaller than that of most normal SNe~Ia. We
note that other SNe~Ia with high expansion velocities, such as SNe
1984A and 2002bo, also exhibit relatively small late-time decline
rates. Thus, slow late-time decline in the blue bands may be common
for rapidly expanding SNe~Ia.

The most notable features in the color curves of SN 2006X are a
plateau in the $U - B$ color after $B$ maximum, and an obvious
deviation from the Lira-Phillips relation at $t = 30$ to 90~d. After
SN 2006X reached its reddest color in $B - V$ at $t \approx 30$~d,
it became progressively bluer at a much faster pace than expected
from the Lira-Phillips relation. We thus suggest that the
Lira-Phillips relation should be applied with caution to any
individual SN~Ia.

We use a global color curve fitting method and the empirical
relations between $\Delta m_{15}$, the peak colors $C_{max}$, and
the post-maximum color $\Delta C_{12}$ to determine the reddening
and extinction of SN 2006X. This yields an average estimate of the
host-galaxy reddening for SN 2006X, $E(B - V)_{host} = 1.42 \pm
0.04$ mag. Assuming the absorption by dust follows the analytical
model proposed by Cardelli et al. (1989), we obtain $R_{V} = 1.48
\pm 0.06$. The low $R_{V}$ value suggests that the dust around SN
2006X is quite different from that observed in the Milky Way Galaxy,
perhaps of a different origin than normal ISM dust. This is also
demonstrated by the relatively smaller EW of the interstellar Na~I~D
absorption in the spectra. Further evidence against conventional
interstellar dust for SN 2006X is that the polarized spectrum of SN
2006X is significantly different from that of extinguished Galactic
stars (Wang L et al. 2006c). It is worth pointing out that most
highly reddened known SNe~Ia, with $E(B - V)_{host} > 0.5$ mag, tend
to have $R_{V}$ values smaller than 3.1. This suggests that the dust
affecting some SNe~Ia may be quite different from that observed in
the Galaxy.

We caution, however, that SN 2006X may have a peculiar (unknown)
intrinsic color evolution, as suggested by the deviation from the
Lira-Phillips relation in the $B - V$ color. If so, it would be
difficult to estimate the host-galaxy reddening toward SN 2006X and
to study its photometric properties.

Spectra of SN 2006X reveal high expansion velocities, based on the
IMEs and the iron-group elements. SN 2006X evolves in a manner
similar to that of other rapidly expanding events such as SNe 1984A
and 2002bo. At early times, these high-$v_{exp}$ objects have much
stronger Si~II $\lambda$6355, Ca~II H\&K, and Ca~II near-IR triplet
lines than other SNe~Ia. At late times, spectra of SN 2006X show a
relatively bluer overall continuum, especially at $\lambda <
4500$~\AA, than the comparison SNe. This is consistent with the
bluer $U - B$ and $B - V$ colors of SN 2006X than the other SNe~Ia
at similar epochs, and suggests an additional input of energy,
probably a light echo.

To explain the high-velocity features of some SNe~Ia, Benetti et al.
(2004) proposed a scenario in which the burning to IMEs extends
farther out into the outermost layers than in normal SNe~Ia. This
model produces the IMEs at higher velocities, but with no additional
$^{56}$Ni production, which provides a plausible explanation for
most of the spectral features seen in SN 2006X. The above scenario
should account not only for the high-velocity feature of the IMEs,
but also for that of the iron-group elements at earlier phases of SN
evolution.

On the other hand, we propose that SN 2006X may be produced in a
progenitor environment having abundant CSM. In this scenario, the
high-velocity features shown in the spectra are formed due to the
density increase caused by interaction between the supernova ejecta
and the surrounding material of the progenitor system, which could
be an accretion disk, a filled Roche lobe, or a
common envelope (Gerardy et~al. 2004). 
The CSM dust, quite possibly associated with the CSM produced by the
progenitor system (Patat et al. 2007), may be the source of a local
light echo that could account for the late-time extra energy (beyond
radioactive decay) seen at short wavelengths. SN 2006X-like events
may actually represent a subclass of SNe Ia with distinct features
of spectroscopy, photometry, polarization and even extinction, and
the presence of significant CSM dust may be one of the possible
explanations (Wang et al. 2007 in preparations). There is also
possibility that the detection of CSM in SN 2006X is just tied to
the particular viewing angle. Obviously, much further research, both
observational and theoretical, is needed on this problem.

\acknowledgments
Some of the data presented here were obtained at
the W. M. Keck Observatory, which is operated as a scientific
partnership among the California Institute of Technology, the
University of California, and the National Aeronautics and Space
Administration (NASA). The Observatory was made possible by the
generous financial support of the W. M. Keck Foundation. We thank
the BAO, Keck, and Lick Observatory staffs for their assistance with
the observations. This research was supported by NSF grant
AST--0607485, the TABASGO Foundation, and the National Natural
Science Foundation of China (NSFC grant 10673007) and the Basic
Research Funding at Tsinghua University (JCqn2005036). KAIT was made
possible by generous donations from Sun Microsystems, Inc., the
Hewlett-Packard Company, AutoScope Corporation, Lick Observatory,
the University of California, and the Sylvia \& Jim Katzman
Foundation. The CTIO 1.3~m telescope is operated by the Smart and
Moderate Aperture Research Telescope System (SMARTS) Consortium. We
are particularly grateful for the scheduling flexibility of SMARTS.
A.V.F. thanks the Aspen Center for Physics, where he participated in
the workshop on ``Supernovae as Cosmological Distance Indicators''
while this paper was nearing completion. We made use of the
NASA/IPAC Extragalactic Database (NED), which is operated by the Jet
Propulsion Laboratory, California Institute of Technology, under
contract with NASA.

\clearpage

\begin{table}
\begin{center}
\caption{Color Terms for Different Telescopes.}
\begin{tabular}{lccccc}
\tableline\tableline Telescopes& $U$ & $B$ & $V$ & $R$ & $I$ \\
\tableline
TNT   & 0.125$\pm$0.016 & $-$0.163$\pm$0.016 & 0.078$\pm$0.007 &  0.098$\pm$0.011 &  $-$0.038$\pm$0.003  \\
KAIT  & \nodata  &$-$0.053$\pm$0.013  & 0.036$\pm$0.007  &  0.069$\pm$0.009 &  $-$0.001$\pm$0.007   \\
CTIO 1.3~m& \nodata & 0.050$\pm$0.002 & $-$0.046$\pm$0.002  &  $-$0.019$\pm$0.004 &$-$0.088$\pm$0.001\\
\tableline
\end{tabular}
\end{center}
\end{table}

\begin{table}
\begin{center}
\caption{Magnitudes of the photometric standards in the field of SN
2006X\tablenotemark{a}} {\scriptsize
\begin{tabular}{lccccccccc}
\tableline\tableline
 Star & $\alpha$(J2000) & $\delta$(J2000) & $U$ & $B$ & $V$ & $R$ & $I$ & $J$ & $K$ \\
\tableline 1
&12$^h$23$^m$03.77$^s$&15$^\circ$47$'$31.6$''$&14.91(02)&14.758(007)
&
14.123(003)&  13.759(005) & 13.393(006) & \nodata &\nodata \\
2    &12:23:10.68&15:44:39.3&17.20(15)&17.069(036) & 16.809(019) &
16.549(027) &
16.177(024) & \nodata & \nodata \\
3    &12:23:05.93&15:43:51.4&16.36(08)&16.613(022) & 16.082(010) &
15.746(014) &
15.385(012) & \nodata & \nodata \\
4    &12:22:48.29&15:43:24.9&\nodata&15.413(013)&  14.763(013) &
14.411(004)     &
14.083(004)   & \nodata & \nodata\\
5    &12:22:42.60&15:49:03.4&14.94(02)&14.753(011) & 14.100(013) &
13.746(014) &
13.404(010)& \nodata & \nodata\\
6    &12:22:50.09&15:50:47.9&16.51(09) & 16.323(014)&15.675(026)&
15.305(028) &
14.941(023)& \nodata & \nodata\\
7    &12:22:51.46&15:51:04.5&16.75(09)& 16.340(020)  &15.458(021) &
14.945(008)&14.445(008) & \nodata & \nodata \\
8    &12:22:53.40&15:52:18.7&17.92(23) &16.380(010)&15.224(022)
&14.425(018)
&13.715(033)& \nodata & \nodata \\
9    &12:22:41.05&15:51:57.8&16.67(09) &16.665(008) &16.131(010)&
15.775(013)
&15.430(005)& \nodata & \nodata \\
10   &12:22:53.98&15:48:40.9&\nodata     &18.233(017) &17.053(012)
&16.309(006)
&15.613(014)&14.721(006)&13.925(009)\\
\tableline

\tablenotetext{a}{See Figure 1 for a chart of SN 2006X (in M100) and
the comparison stars.}
\end{tabular}
}
\end{center}
\end{table}


\clearpage
\setlength{\hoffset}{-12mm}
\begin{deluxetable}{lccccccccl}
\tablewidth{560pt} \tabletypesize{\scriptsize} \tablecaption{Optical
photometry of SN 2006X.}
\tablehead{
 \colhead{UT Date} &
 \colhead{JD$-$2,450,000} &
 \colhead{Phase\tablenotemark{*}}&
 \colhead{$U$} &
 \colhead{$B$} &
 \colhead{$V$} &
 \colhead{$R$} &
 \colhead{$I$} &
 \colhead{Method}&
 \colhead{Telescope}
 }
\startdata
\multicolumn{10}{c}{}\\
\noalign{\smallskip}
 02/08/2006 & 3774.90 &   $-$11.27  &  \nodata       & 17.111(0.023) & 15.679(0.019) & 14.832(0.019) & 14.745(0.020) &PSF & KAIT\\
 02/08/2006 & 3775.33 &   $-$10.84  &  17.576(0.048) & 17.005(0.018) & 15.566(0.014) & 14.721(0.013) & 14.489(0.017) &PSF &TNT \\
 02/09/2006 & 3775.85 &   $-$10.32  &  \nodata       & 16.701(0.020) & 15.421(0.019) & 14.574(0.015) & 14.416(0.021) &PSF & KAIT \\
 02/09/2006 & 3776.19 &   $-$9.98  &  17.449(0.153) & 16.626(0.019)  & 15.320(0.014)  & 14.476(0.013) & 14.318(0.017) &PSF & TNT \\
 02/10/2006 & 3776.86 &   $-$9.31   &  \nodata       & 16.484(0.021) & 15.146(0.020)& \nodata       & \nodata       &APER & CTI0 1.3m  \\
 02/10/2006 & 3776.96 &   $-$9.21   &  \nodata       & 16.386(0.054) & 15.154(0.018)& 14.327(0.021) & 14.150(0.024) &PSF& KAIT  \\
 02/10/2006 & 3777.38 &   $-$8.79   &  17.485(0.159) & 16.297(0.015) & 15.012(0.014)& 14.207(0.012) & 13.925(0.016) &PSF& TNT \\
 02/11/2006 & 3778.38 &   $-$7.79   &  16.644(0.106) & 16.060(0.018) & 14.797(0.015)& 14.032(0.013) & 13.731(0.017) &PSF& TNT \\
 02/12/2006 & 3778.75 &   $-$7.42   &  \nodata       & 15.994(0.019) & 14.813(0.018)& 14.043(0.014) & 13.781(0.022) &PSF&  KAIT  \\
 02/13/2006 & 3779.80 &   $-$6.37   &  \nodata       & 15.778(0.020) & 14.545(0.020)& \nodata       & \nodata       &APER& CTIO 1.3m \\
 02/13/2006 & 3780.39 &   $-$5.78   &  \nodata       & 15.741(0.017) & 14.469(0.015)& 13.786(0.013) & 13.443(0.017) &PSF& TNT \\
 02/15/2006 & 3782.40 &   $-$3.77   &  \nodata       & 15.633(0.041) & 14.343(0.017)& 13.660(0.013) & 13.300(0.019) &PSF& TNT \\
 02/16/2006 & 3783.80 &   $-$2.37   &  \nodata       & 15.389(0.020) & 14.166(0.020)& 13.619(0.015) & 13.298(0.015) &APER& CTIO 1.3m \\
 02/17/2006 & 3784.27 &   $-$1.90   &  \nodata       & 15.441(0.014) & 14.146(0.014)& 13.552(0.013) & \nodata       &PSF & TNT \\
 02/18/2006 & 3784.76 &   $-$1.51   &  \nodata       & 15.406(0.020) & 14.178(0.020)& 13.586(0.015) & 13.297(0.015) &APER& CTIO 1.3m \\
 02/18/2006 & 3785.25 &   $-$0.92   & 16.256(0.034)  & 15.431(0.015) & 14.094(0.014)& 13.521(0.013) & 13.283(0.016) &PSF& TNT \\
 02/19/2006 & 3786.19 &    0.02   & 16.143(0.029)  & 15.407(0.014) & 14.059(0.014) &13.499(0.012) & 13.290(0.016) &PSF& TNT \\
 02/21/2006 & 3787.78 &    1.61   &  \nodata       & 15.418(0.028) & 14.070(0.023) &13.518(0.015) & 13.354(0.015) &APER& CTIO 1.3m \\
 02/21/2006 & 3787.84 &    1.67   &  \nodata       & 15.466(0.015) & 14.068(0.017) &13.524(0.013) & 13.409(0.018) &PSF&  KAIT  \\
 02/21/2006 & 3788.27 &    2.10   & 16.273(0.022)  & 15.441(0.014) & 14.024(0.014) &13.477(0.012) & 13.323(0.016) &PSF& TNT \\
 02/22/2006 & 3788.82 &    2.65   &  \nodata       & 15.494(0.015) & 14.054(0.020) &13.513(0.014) & 13.412(0.022) &PSF&  KAIT  \\
 02/22/2006 & 3789.25 &    3.08   & 16.333(0.024)  & 15.499(0.014) & 14.013(0.014) &13.476(0.013) & 13.338(0.016) &PSF& TNT \\
 02/23/2006 & 3789.85 &    3.68   &  \nodata       & 15.553(0.015) & 14.063(0.017) &13.526(0.015) & 13.444(0.018) &PSF &  KAIT  \\
 02/23/2006 & 3790.23 &    4.06   & 16.401(0.040)  & 15.542(0.014) & 14.025(0.014) &13.507(0.012) & 13.376(0.016) &PSF & TNT \\
 02/24/2006 & 3790.78 &    4.61   &  \nodata       & 15.585(0.032) & 14.111(0.020) &13.570(0.015) & 13.468(0.015) &APER& CTIO 1.3m \\
 02/24/2006 & 3790.83 &    4.66   &  \nodata       & 15.618(0.015) & 14.062(0.014) &13.543(0.041) & 13.496(0.030) &PSF&  KAIT  \\
 02/25/2006 & 3792.23 &    6.06   & 16.545(0.034)  & 15.685(0.014) & 14.063(0.014) &13.579(0.012) & 13.461(0.016) &PSF& TNT \\
 02/26/2006 & 3793.23 &    7.06   & 16.594(0.083)  & 15.771(0.016) & 14.085(0.014) &13.620(0.013) & 13.508(0.017) &PSF& TNT \\
 02/27/2006 & 3793.83 &    7.66   &  \nodata       & 15.805(0.020) & 14.126(0.018) &13.712(0.015) & 13.645(0.015) &APER& CTIO 1.3m \\
 02/28/2006 & 3795.34 &    9.17   & 16.794(0.036)  & 15.943(0.015) & 14.186(0.014) &13.784(0.012) & 13.643(0.016) &PSF& TNT \\
 03/02/2006 & 3797.39 &   11.22   & 17.003(0.049)  & 16.136(0.015) & 14.335(0.014) &13.949(0.013) & 13.749(0.017) &PSF& TNT \\
 03/04/2006 & 3799.39 &   13.22   & 17.246(0.047)  & 16.338(0.015) & 14.428(0.014) &14.040(0.013) & 13.757(0.017) &PSF& TNT \\
 03/05/2006 & 3799.74 &   13.57   &  \nodata       & 16.473(0.031) & 14.488(0.041) &14.048(0.015) & 13.938(0.023) &APER& CTIO 1.3m \\
 03/05/2006 & 3800.32 &   14.15   & 17.379(0.079)  & 16.462(0.016) & 14.516(0.014) &14.078(0.012) & 13.742(0.016) &PSF& TNT \\
 03/06/2006 & 3801.22 &   15.05   & 17.583(0.117)  & 16.562(0.016) & 14.562(0.014) &14.083(0.012) & 13.729(0.016) &PSF& TNT \\
 03/07/2006 & 3802.23 &   16.06   &  \nodata       & 16.654(0.019) & 14.610(0.015) &14.082(0.013) &   \nodata     &PSF& TNT \\
 03/08/2006 & 2802.73 &   16.56   &  \nodata       & 16.790(0.020) & 14.682(0.017) &14.103(0.015) & 13.864(0.015) &APER& CTIO 1.3m \\
 03/09/2006 & 3803.22 &   17.05   &  \nodata       & 16.760(0.018) & 14.631(0.014) &14.084(0.013) & 13.661(0.017) &PSF& TNT \\
 03/11/2006 & 3806.16 &   19.99   &  \nodata       & 17.240(0.045) & 14.777(0.016) &14.096(0.013) & 13.603(0.017) &PSF & TNT \\
 03/12/2006 & 3806.74 &   20.57   &  \nodata       & 17.303(0.030) & 14.899(0.020) &14.138(0.015) & 13.740(0.015) &APER & CTIO 1.3m \\
 03/13/2006 & 3808.25 &   22.08   &  \nodata       & 17.243(0.095) & 14.821(0.017) &  \nodata     & 13.575(0.018) &PSF& TNT \\
 03/15/2006 & 3810.31 &   24.14   &  \nodata       & 17.506(0.104) & 14.930(0.020) &14.135(0.015) &   \nodata     &PSF& TNT \\
 03/16/2006 & 3810.70 &   24.53   &  \nodata       & 17.743(0.108) & 15.045(0.018) &14.200(0.015) & 13.650(0.015) &APER& CTIO 1.3m \\
 03/18/2006 & 3813.23 &   27.06   & 18.959(0.343)  & 17.697(0.024) & 15.120(0.014) &14.214(0.012) & 13.518(0.016) &PSF& TNT \\
 03/19/2006 & 3813.82 &   27.65   &  \nodata       & 17.729(0.042) & 15.221(0.017) &14.313(0.014) & 13.603(0.018) &PSF& KAIT \\
 03/20/2006 & 3814.71 &   28.54   &  \nodata       & 17.945(0.054) & 15.388(0.015) &14.433(0.015) & 13.709(0.015) &APER& CTIO 1.3m \\
 03/21/2006 & 3815.24 &   29.07   &  \nodata       & 17.884(0.022) & 15.245(0.014) &14.350(0.013) & 13.566(0.018) &PSF& TNT \\
 03/22/2006 & 3816.80 &   30.63   &  \nodata       & 17.878(0.039) & 15.454(0.016) &14.538(0.015) & 13.792(0.025) &PSF&  KAIT  \\
 03/22/2006 & 3817.19 &   31.02   & 19.017(0.212)  & 17.948(0.018) & 15.432(0.014) &14.510(0.012) & 13.741(0.016) &PSF& TNT \\
 03/23/2006 & 3818.15 &   31.98   &  \nodata       & 18.030(0.023) & 15.509(0.014) &14.588(0.013) & 13.818(0.016) &PSF& TNT \\
 03/24/2006 & 3818.72 &   32.55   &  \nodata       & 18.091(0.033) & 15.607(0.033) &14.722(0.015) & 13.976(0.015) &APER& CTIO 1.3m \\
 03/25/2006 & 3820.17 &   34.00   & 19.087(0.226)  & 18.166(0.020) & 15.631(0.014) &14.735(0.012) & 13.976(0.016) &PSF& TNT \\
 03/27/2006 & 3821.77 &   35.60   & \nodata        & \nodata       & 15.749(0.017) &14.871(0.030) & 14.170(0.043) &PSF&  KAIT  \\
 03/28/2006 & 3823.20 &   37.00   & 19.055(0.373)  & 18.132(0.035) & 15.769(0.015) &14.884(0.013) & 14.156(0.016) &PSF& TNT \\
 03/29/2006 & 3824.16 &   37.99   & 19.334(0.326)  & 18.188(0.029) & 15.814(0.015) &14.949(0.013) & 14.215(0.016) &PSF& TNT \\
 03/30/2006 & 3824.77 &   38.60   & \nodata        & 18.206(0.030) & 15.923(0.042) &15.116(0.015) & 14.371(0.011) &APER& CTIO 1.3m\\
 03/30/2006 & 3825.10 &   38.93   &  \nodata       & 18.278(0.052) & 15.839(0.017) &14.987(0.013) & 14.265(0.017) &PSF& TNT \\
 04/01/2006 & 3827.18 &   41.01   &  \nodata       & 18.306(0.043) & 15.885(0.016) &15.091(0.013) & 14.377(0.017) &PSF& TNT \\
 04/02/2006 & 3828.20 &   42.02   & 19.368(0.344)  & 18.261(0.028) & 15.922(0.015) &15.088(0.013) & 14.429(0.017) &PSF& TNT \\
 04/03/2006 & 3829.25 &   43.08   &  \nodata       & 18.210(0.110) & 15.984(0.034) &15.153(0.014) & 14.475(0.018) &PSF& TNT \\
 04/04/2006 & 3830.15 &   43.98   &  \nodata       & 18.287(0.027) & 15.997(0.015) &15.171(0.013) & 14.516(0.016) &PSF& TNT \\
 04/07/2006 & 3833.24 &   47.07   &  \nodata       & 18.315(0.043) & 16.080(0.017) &15.285(0.013) & 14.651(0.017) &PSF& TNT \\
 04/11/2006 & 3836.60 &   50.43   &  \nodata       & 18.471(0.086) & 16.288(0.025) &15.445(0.022) & 14.913(0.053) &APER& CTIO 1.3m \\
 04/12/2006 & 3838.14 &   51.97   &  \nodata       & \nodata       &  \nodata      &15.382(0.037) & 14.824(0.027) &PSF& TNT \\
 04/16/2006 & 3841.65 &   55.48   &  \nodata       & 18.523(0.053) & 16.440(0.014) &15.689(0.013) & 15.176(0.010) &APER& CTIO 1.3m\\
 04/18/2006 & 3843.72 &   57.55   &  \nodata       & 18.443(0.054) & 16.386(0.019) &15.670(0.015) & 15.173(0.028) &PSF&  KAIT  \\
 04/19/2006 & 3845.13 &   58.86   &  \nodata       & 18.487(0.023) & 16.440(0.015) &15.695(0.013) & 15.171(0.017) &PSF& TNT \\
 04/21/2006 & 3846.63 &   60.46   &  \nodata       & 18.444(0.032) & 16.541(0.014) &15.771(0.013) & 15.299(0.015) &APER& CTIO 1.3m \\
 04/24/2006 & 3850.19 &   64.02   &  \nodata       & 18.492(0.062) & 16.538(0.023) &15.826(0.015) & 15.300(0.022) &PSF& TNT \\
 04/27/2006 & 3852.65 &   66.48   &  \nodata       & 18.519(0.041) & 16.699(0.019) &15.947(0.015) & 15.576(0.018) &APER& CTIO 1.3m \\
 04/27/2006 & 3852.68 &   66.51   &  \nodata       & 18.506(0.070) & 16.645(0.018) &15.949(0.013) & 15.514(0.021) &PSF&  KAIT  \\
 04/28/2006 & 3854.08 &   67.91   &  \nodata       & 18.555(0.038) & 16.670(0.018) &16.015(0.014) & 15.561(0.018) &PSF& TNT \\
 05/03/2006 & 3858.62 &   72.45   &  \nodata       & 18.599(0.040) & 16.868(0.033) & \nodata       & \nodata       &APER& CTIO 1.3m \\
 05/03/2006 & 3859.04 &   72.87   &  \nodata       & 18.598(0.064) & 16.780(0.025) & 16.139(0.016) & 15.748(0.020) &PSF& TNT \\
 05/05/2006 & 3864.66 &   78.49   &  \nodata       & 18.818(0.105) & 16.962(0.054) & \nodata       & \nodata       &APER& CTIO 1.3m \\
 05/10/2006 & 3866.07 &   79.90   &  \nodata       & 18.589(0.229) & 16.931(0.053) & 16.368(0.023) & 15.978(0.024) &PSF& TNT \\
 05/14/2006 & 3870.07 &   83.90   &  \nodata       & 18.622(0.061) & 17.091(0.024) & 16.488(0.016) & 16.096(0.020) &PSF& TNT \\
 05/18/2006 & 3873.58 &   87.41   &  \nodata       & 18.662(0.023) & 17.249(0.011) & \nodata       & \nodata       &APER& CTIO 1.3m\\
 05/23/2006 & 3879.16 &   92.99   &  \nodata       & 18.719(0.147) & 17.314(0.043) & 16.721(0.026) & 16.494(0.040) &PSF& TNT \\
 05/26/2006 & 3881.60 &   95.43   &  \nodata       & 18.773(0.020) & 17.419(0.010) & \nodata       & \nodata       &APER& CTIO 1.3m \\
 05/28/2006 & 3884.16 &   97.99   &  \nodata       & 18.795(0.031) & 17.467(0.020) &16.982(0.016) & 16.652(0.020) &PSF& TNT \\
 06/07/2006 & 3894.08 &   107.91  &  \nodata       & 19.025(0.179) & 17.643(0.032) & 17.279(0.023) & 16.878(0.041) &PSF& TNT \\
 06/14/2006 & 3901.08 &   114.91  &  \nodata       & 18.886(0.054) & 17.789(0.029) & 17.462(0.023) & 17.064(0.045) &PSF& TNT \\
 06/16/2006 & 3903.03 &   116.86  &  \nodata       & 19.033(0.199) & 17.969(0.028) & 17.527(0.016) & 17.383(0.021) &PSF& TNT \\
\enddata
\tablenotetext{*}{Relative to the epoch of $B$-band maximum (JD =
2,453,786.17).}
\end{deluxetable}

\clearpage
\setlength{\hoffset}{0mm}

\begin{table}
\begin{center}
\caption{Infrared photometry of SN 2006X.}
\begin{tabular}{cccccl}
\tableline
UT Date & JD$-$2,450,000 &Phase\tablenotemark{*} & $J$ & $K$  \\
\tableline
02/10/2006 &3776.86 & $-$9.31 & 13.353(0.013) &13.178(0.038) \\
02/13/2006 &3779.80 & $-$6.37 & 12.969(0.015) &12.885(0.034) \\
02/17/2006 &3783.80 & $-$2.37 & 12.921(0.015) &12.881(0.030) \\
02/18/2006 &3784.76 & $-$1.41 & 12.957(0.015) &12.854(0.030) \\
02/21/2006 &3787.78 &  1.61 & 13.129(0.013) &12.912(0.045) \\
02/24/2006 &3790.78 &  4.61 & 13.360(0.013) &13.061(0.043)  \\
02/27/2006 &3793.83 &  7.66 & 13.802(0.018) &13.229(0.058) \\
03/02/2006 &3796.75 & 10.58 & 14.433(0.019) &13.239(0.040) \\
03/05/2006 &3799.74 & 13.57 & 14.785(0.029) & \nodata      \\
03/08/2006 &3802.73 & 16.56 & 14.526(0.017) &13.120(0.026)  \\
03/12/2006 &3806.74 & 20.57 & 14.328(0.017) &12.945(0.039)  \\
03/16/2006 &3810.70 & 24.53 & 14.071(0.016) &12.854(0.034)  \\
03/20/2006 &3814.71 & 28.54 & 13.780(0.016) &12.899(0.051)  \\
03/24/2006 &3818.72 & 32.55 & 13.909(0.018) &12.933(0.043)  \\
03/30/2006 &3824.77 & 38.60 & 14.480(0.019) &13.648(0.028) \\
04/11/2006 &3836.60 & 50.43 & 15.282(0.027) &14.103(0.073)  \\
04/16/2006 &3841.65 & 55.48 & 15.727(0.034) &14.350(0.043)  \\
04/21/2006 &3846.63 & 60.46 & 15.943(0.036) &14.569(0.043)  \\
04/27/2006 &3852.65 & 66.48 & 16.302(0.047) &14.601(0.058)  \\
05/03/2006 &3858.62 & 72.45 & 16.521(0.128) &  \nodata      \\
05/09/2006 &3864.66 & 78.49 & 17.121(0.096) &15.124(0.072)  \\
\tableline
\end{tabular} \tablenotetext{*}{Relative to the epoch of $B$-band
maximum (JD = 2,453,786.17).}
\end{center}
\end{table}


\begin{table}
\begin{center}
\caption{Optical and NIR S-corrections added to the data of CTIO
1.3~m telescope listed in Tables 3 and 4.}
\begin{tabular}{ccrrrrr}
\tableline
UT Date & JD$-$2,450,000 & Phase\tablenotemark{*} & B &  V &  J &  K \\
\tableline
02/10/2006 & 3776.86 & $-$9.31  & $-$0.046 & $-$0.004 & 0.038  & $-$0.003  \\
02/13/2006 & 3779.80 & $-$6.37  & $-$0.043 & 0.001  & 0.046    & 0.005   \\
02/17/2006 & 3783.80 & $-$2.37  & $-$0.043 & 0.010   & 0.059   & 0.019   \\
02/18/2006 & 3784.76 & $-$1.41  & $-$0.044 & 0.012  & 0.062    & 0.022   \\
02/21/2006 & 3787.78 & 1.61   & $-$0.044 & 0.017  & 0.067      & 0.034   \\
02/24/2006 & 3790.78 & 4.61   & $-$0.043 & 0.021  & 0.036     & 0.057   \\
02/27/2006 & 3793.83 & 7.66   & $-$0.039 & 0.023  & 0.007    & 0.054   \\
03/02/2006 & 3796.75 & 10.58  & \nodata& \nodata & $-$0.018  & 0.043   \\
03/05/2006 & 3799.74 & 13.57  & $-$0.024 & 0.026  & $-$0.043 & \nodata    \\
03/08/2006 & 3802.73 & 16.56  & $-$0.014 & 0.027  & $-$0.066 & 0.018   \\
03/12/2006 & 3806.74 & 20.57  & $-$0.003 & 0.029  & $-$0.091 & 0.004   \\
03/16/2006 & 3810.70 & 24.53  & $-$0.019 & 0.032  & $-$0.076 & $-$0.010   \\
03/20/2006 & 3814.71 & 28.54  & $-$0.066 & 0.036  & $-$0.057 & $-$0.016  \\
03/24/2006 & 3818.72 & 32.55  & $-$0.069 & 0.039  & $-$0.09  & $-$0.007  \\
03/30/2006 & 3824.77 & 38.60  & $-$0.066 & 0.041  & $-$0.119 & $-$0.005  \\
04/11/2006 & 3836.60 & 50.43  & $-$0.056 & 0.030  & $-$0.131 & 0   \\
04/16/2006 & 3841.65 & 55.48  & $-$0.049 & 0.025  & $-$0.131 & 0   \\
04/21/2006 & 3846.63 & 60.46  & $-$0.041 & 0.026  & $-$0.131 & 0   \\
04/27/2006 & 3852.65 & 66.48  & $-$0.032 & 0.026  & $-$0.131 & 0   \\
05/03/2006 & 3858.62 & 72.45  & $-$0.032 & 0.025  & $-$0.131 & 0   \\
05/09/2006 & 3864.66 & 78.49  & $-$0.032 & 0.025  & $-$0.131 & 0   \\
05/18/2006 & 3873.58 & 87.41  & $-$0.032 & 0.025  &\nodata & \nodata   \\
05/26/2006 & 3881.60 & 95.43  & $-$0.032 & 0.025  &\nodata & \nodata   \\

\tableline
\end{tabular} \tablenotetext{*}{Relative to the epoch of $B$-band
maximum (JD = 2,453,786.17).}
\end{center}
\end{table}


\begin{table}[htb]
\begin{center}
\caption{Journal of spectroscopic observations of SN 2006X.}
\begin{tabular}{cccccl}
\tableline\tableline
UT Date& JD &Phase\tablenotemark{a}&Range&Resolution\tablenotemark{b}&Instrument\\
 &$-$2,450,000&(days)&\AA& \AA & \\
\tableline
18/02/06&3785.3&$-$0.9&3700-8500&4&BAO 2.16m Cassegrain\\
21/02/06&3787.7&+1.5&3300-11000 &5-12&Lick Shane 3.0m Kast\\
21/02/06&3788.3&+2.1&4000-9000 &3&BAO 2.16m BFOSC \\
25/02/06&3792.3&+6.1&3700-8200  &4&BAO 2.16m Cassegrain\\
04/03/06&3799.2&+13.1&3700-8200&4&BAO 2.16m Cassegrain\\
21/03/06&3816.2&+30.1&4000-9000 &3&BAO 2.16m BFOSC\\
22/03/06&3816.7&+30.6&3300-11000&5-12&Lick Shane 3.0m Kast\\
01/04/06&3827.2&+41.0&4500-6200 &4&BAO 2.16m Cassegrain\\
07/04/06&3833.3&+47.1&4500-9000 &3&BAO 2.16m BFOSC\\
28/04/06&3853.9&+67.7&3900-10500&5-12&Lick Shane 3.0m Kast\\
05/05/06&3860.9&+74.7&3900-10500&6-12&Lick Shane 3.0m Kast\\
28/05/06&3884.0&+97.8&3300-11000&5-11&Lick Shane 3.0m Kast\\
11/23/06&4063.2&+277.0&3100-9200&6&Keck I 10m LRIS \\
12/23/06&4093.2&+307.0&4500-7200&0.35&Keck II 10m DEIMOS \\
\tableline
\end{tabular} \tablenotetext{a}{Relative to the $B$ maximum(JD=2453786.17)}
\tablenotetext{b}{Approximate Spectral Resolution (full width at
half maximum intensity).}
\end{center}
\end{table}

\begin{table}[htb]
\begin{center}
\caption{Light-curve parameters of SN 2006X.} {\scriptsize
\begin{tabular}{lccc}
\tableline\tableline
Band& $t_{max}$ & $m_{peak}$ & $\beta$\tablenotemark{a} \\
\tableline
U &2453785.13$\pm$0.96 &16.20$\pm$0.06& ... \\
B &2453786.17$\pm$0.35 &15.40$\pm$0.03&0.92$\pm$0.05\\
V &2453789.11$\pm$0.29 &14.04$\pm$0.03&2.71$\pm$0.04\\
R &2453789.09$\pm$0.32 &13.50$\pm$0.03&3.30$\pm$0.04\\
I &2453784.73$\pm$0.40 &13.29$\pm$0.10&3.60$\pm$0.02\\
J &2453782.09$\pm$0.38 &12.88$\pm$0.02&6.29$\pm$0.24\\
K &2453782.97$\pm$0.46 &12.82$\pm$0.04&3.78$\pm$0.15\\
\tableline
\end{tabular}}
\tablenotetext{a}{The late-time decline rate (in units of mag
(100~d)$^{-1}$) of the light curve.}
\end{center}
\end{table}


\begin{table}[htb]
\begin{center}
\caption{Host-galaxy reddening of SN 2006X derived from different
methods.}
\begin{tabular}{lcccccc}
\tableline\tableline
Method& $E(U - B)$ & $E(B - V)$ & $E(V - R)$ & $E(V - I)$ & $E(V - J)$ & $E(V - K)$ \\
\tableline Global color
fit&...&1.41$\pm$0.06&0.61$\pm$0.05&1.27$\pm$0.12&1.76$\pm$0.09&1.95$\pm$0.09\\
$C_{max}$ &1.23$\pm$0.13&1.38$\pm$0.07&0.54$\pm$0.07&1.02$\pm$0.13
&1.71$\pm$0.10 &
2.04$\pm$0.09 \\
$C_{12}$   &...          &1.45$\pm$0.06&0.66$\pm$0.06&1.33$\pm$0.12
&1.73$\pm$0.11 &
1.85$\pm$0.11 \\
Mean       &1.23$\pm$0.13&1.42$\pm$0.04&0.61$\pm$0.03&1.22$\pm$0.07
&1.74$\pm$0.06
&1.96$\pm$0.06\\
\tableline
\end{tabular}
\end{center}
\end{table}


\begin{table}[htb]
\begin{center}
\caption{Parameters of  highly extinguished SNe Ia.} {\scriptsize
\begin{tabular}{lccclc}
\tableline\tableline
SN&Host Galaxy & $E(B - V)_{host}$ & $R_{V}$ & Colors & Source\\
\tableline
1995E &NGC 4419&0.74$\pm$0.05&2.62$\pm$0.28&$BVI$  &1,2\\
1996ai&NGC 5005&1.69$\pm$0.10&2.03$\pm$0.12&$BVI$  &1,2\\
1999cl&NGC 4501&1.24$\pm$0.07&1.55$\pm$0.08&$BVRIJHK$&3\\
2000ce&UGC 4195&0.60$\pm$0.06&2.70$\pm$0.30&$UBVRI$&1,2\\
2003cg&NGC 3169&1.33$\pm$0.11&1.80$\pm$0.19&$UBVRIJHK$&4 \\
2006X &NGC 4321&1.42$\pm$0.06&1.50$\pm$0.15&$UBVRIJK$& This paper\\
 \tableline
\end{tabular}}
\tablerefs{(1)Wang et~al. (2006); (2) Jha et~al.
(2007); (3) Krisciunas et~al. (2006); (4) Elias-Rosa et~al. (2006).}
\end{center}
\end{table}

\begin{table}[htb]
\begin{center}
\caption{Absolute magnitudes of SN 2006X at maximum light.}
{\scriptsize
\begin{tabular}{lccccccc}
\tableline\tableline
Method& $M_{U}$ & $M_{B}$ & $M_{V}$ & $M_{R}$ & $M_{I}$ & $M_{J}$ & $M_{K}$\\
\tableline Cepheid distance
&$-$19.63$\pm$0.25&$-$19.10$\pm$0.20&$-$19.06$\pm$0.17&$-$19.02$\pm$0.15&$-$18.52$\pm$0.18
&$-$18.37$\pm$0.14 &$-$18.23$\pm$0.14\\
Normalized to $\Delta m_{15}=1.10$& \nodata
&$-$19.23$\pm$0.26&$-$19.18$\pm$0.22&$-$19.13$\pm$0.21&$-$18.62$\pm$0.23&
\nodata &
\nodata  \\
Normalized to $\Delta C_{12}=0.31$
&$-$19.92$\pm$0.30&$-$19.31$\pm$0.23&$-$19.22$\pm$0.21&\nodata&$-$18.63$\pm$0.21
&
\nodata  &\nodata\\
\tableline
\end{tabular}}
\end{center}
\end{table}
\begin{table}[htb]
\begin{center}
\caption{Relevant parameters for SN 2006X and its host galaxy.}
{\scriptsize
\begin{tabular}{lcc}
\tableline\tableline
& SN 2006X photometric parameters& \\
\tableline
UT discovery date& 7.15 February 2006 & 1\\
Epoch of $B$ maximum & 2,453,786.17$\pm$0.35 & 2\\
$B_{max}$  & 15.40$\pm$0.03 & 2 \\
$B_{max} - V_{max}$ & 1.36$\pm$0.04 & 2\\
$E(B - V)_{host}$ & 1.42$\pm$0.04 & 2\\
$R_{V}$ & 1.48$\pm$0.06 & 2\\
$M^{B}_{max}$ & $-$19.10$\pm$0.20 & 2 \\
$\Delta m_{15}$(observed)&1.17$\pm$0.05 & 2\\
$\Delta m_{15}$(true) & 1.31$\pm$0.05 & 2\\
$\Delta C_{12}$(observed) & 1.83$\pm$0.05 & 2 \\
$\Delta C_{12}$(true) &0.42$\pm$0.06 & 2\\
Late-time $B$ decline rate & 0.92$\pm$0.05 mag (100~d)$^{-1}$ & 2\\
$H_{0}$ & 72.8$\pm$8.2 km s$^{-1}$ Mpc$^{-1}$ & 2\\
L$^{max}_{bol}$ & 1.02$\pm$0.10 $\times 10^{43}$ erg s$^{-1}$ & 2\\
$t_{r}$ & 18.2$\pm$0.9 day & 2 \\
M($^{56}$Ni) & 0.50$\pm$0.05~M$_{\odot}$ & 2\\
\tableline\tableline
& SN 2006X spectroscopic parameters & \\
\tableline
$v_{max}$(Si II $\lambda$6355) & $\sim$15500 km s$^{-1}$ & 2\\
$v_{max}$(Ca II H\&K) & $\sim$20500km s$^{-1}$  & 2\\
$v_{max}$ (S II $\lambda$5460) & $\sim$11800 km s$^{-1}$ &2 \\
$v_{30}$ (Fe II $\lambda$4555) &$\sim$10300 km s$^{-1}$ & 2 \\
$\dot{v}$ (Si II $\lambda$6355)& 123$\pm$10 km s$^{-1}$ &2 \\
R(Si II)  & 0.12$\pm$0.06 & 2\\
 \tableline
 \tableline
 & Parameters for NGC 4321 & \\
 \tableline
 Galaxy type & Sbc, LINER/ H II & 3\\
 $E(B - V)_{Gal}$ & 0.026& 3\\
 $(m-M)_{Cepheid}$ & 30.91$\pm$0.14 & 4 \\
 $v_{hel}$ & 1557 km s$^{-1}$ & 3\\
 \tableline

\end{tabular}}
\tablerefs{(1) Suzuki \& Migliardi (2006); (2)
this paper; (3) NASA Extragalactic Database; (4) Freedman et~al.
(2001)}
\end{center}
\end{table}

\clearpage

\begin{figure}
\figurenum{1}
\includegraphics[angle=0,width=160mm]{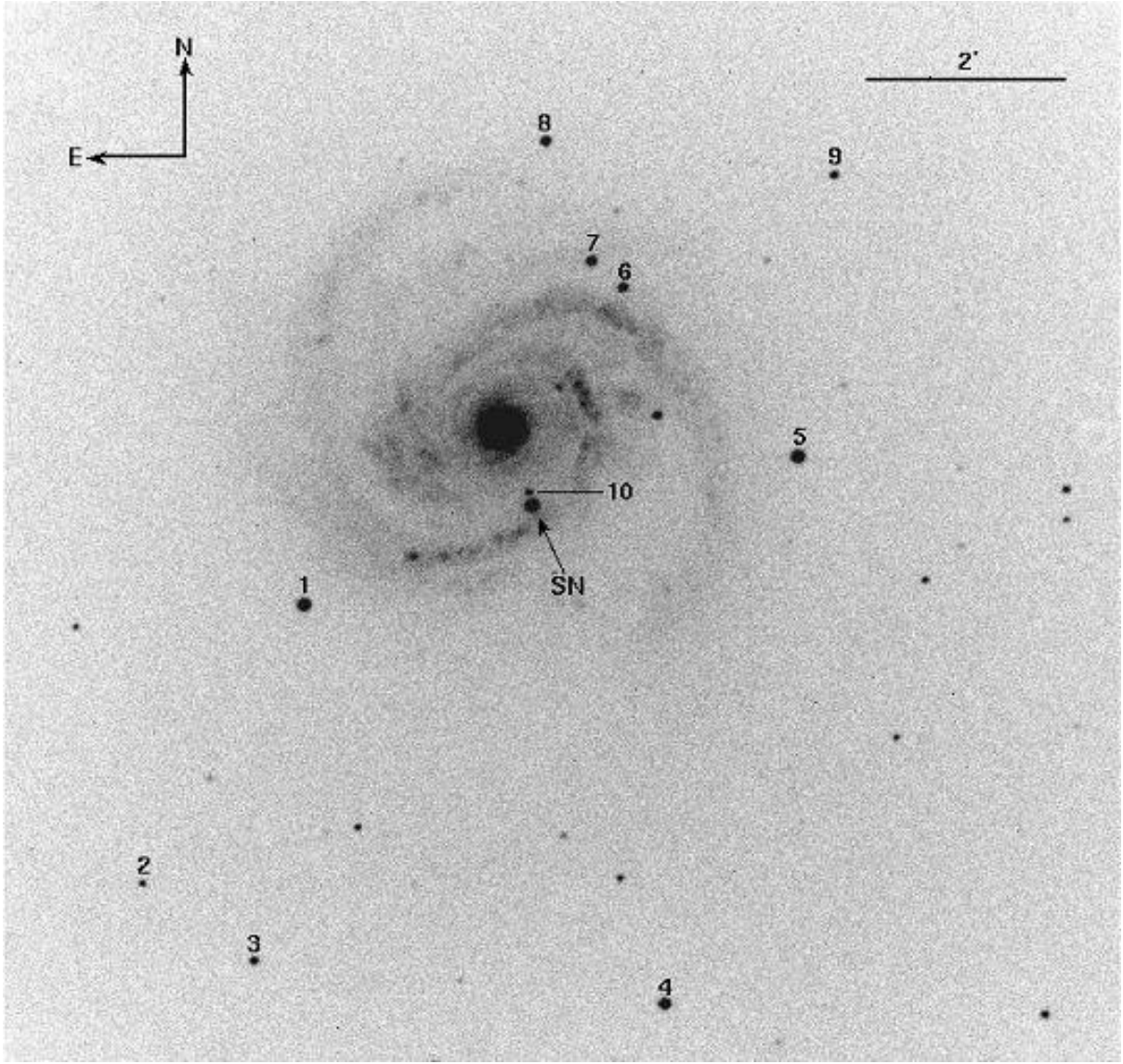}
\caption{SN 2006X in NGC 4321 (M100). This is a $V$-band image taken
by the 0.8~m TNT on 19.69 Feb. 2006. The supernova and 10 local
reference stars (Table 2) are marked by numbers.}\label{fig-1}
\vspace{-0.0cm}
\end{figure}

\begin{figure}
\figurenum{2} \vspace*{-40mm}\plotone {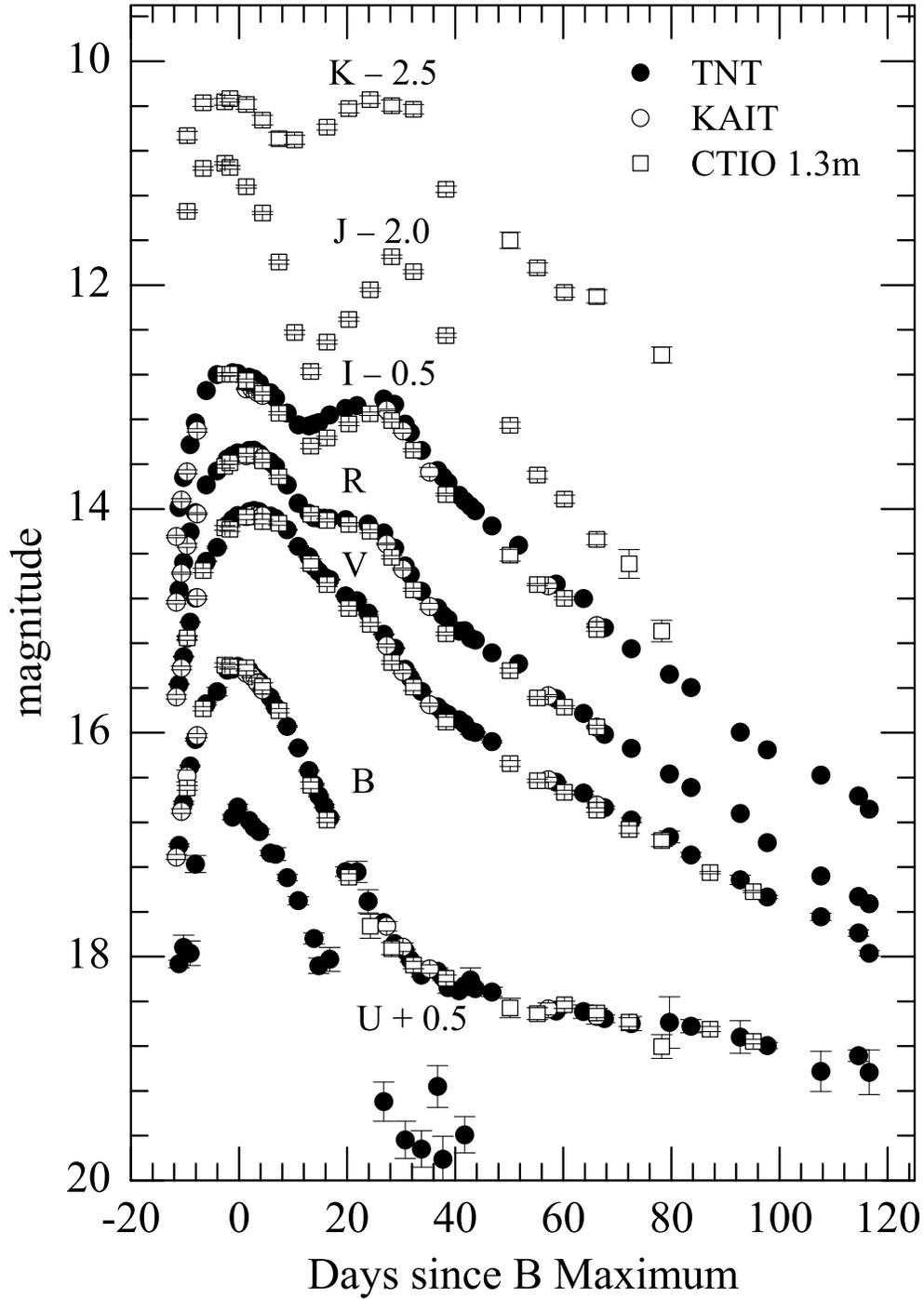} \vspace{-2.5cm} \hspace{-0.2cm}
 \caption{The $UBVRIJK$ light curves of SN 2006X. Solid dots denote the
$UBVRI$ observations from TNT; open circles show the $BVRI$ data
from KAIT; open squares represent the $BVRIJK$ data from the CTIO
1.3~m telescope. The $U$, $I$, $J$, and $K$ data have been offset
vertically by +0.5, $-$0.5, $-$2.0, and $-$2.5 mag, respectively.}
\label{fig-2} \vspace{-0.2cm}
\end{figure}

\begin{figure}
\figurenum{3} \plotone {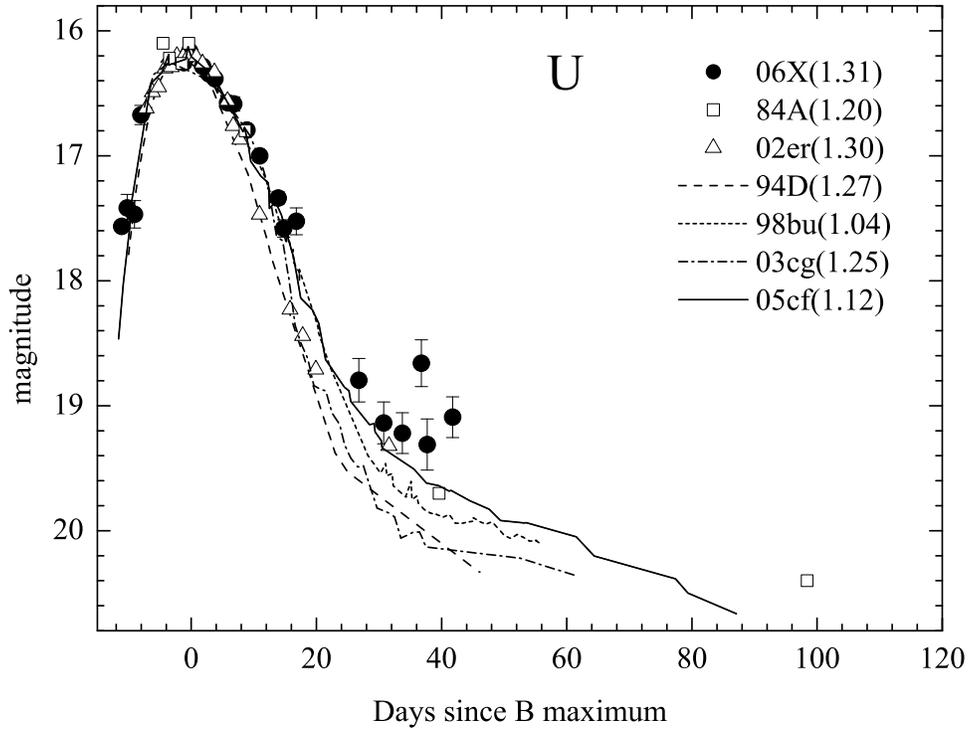}
 \caption{Comparison of the observed $U$-band light curves of
SN 2006X with those of SNe 1984A, 1994D, 1998bu, 2002er, 2003cg, and
2005cf. All light curves are arbitrarily shifted in time and
magnitude to fit the peak value of SN 2006X. The data sources are
cited in the text.}
 \label{fig-3}
\end{figure}

\begin{figure}
\figurenum{4} \plotone {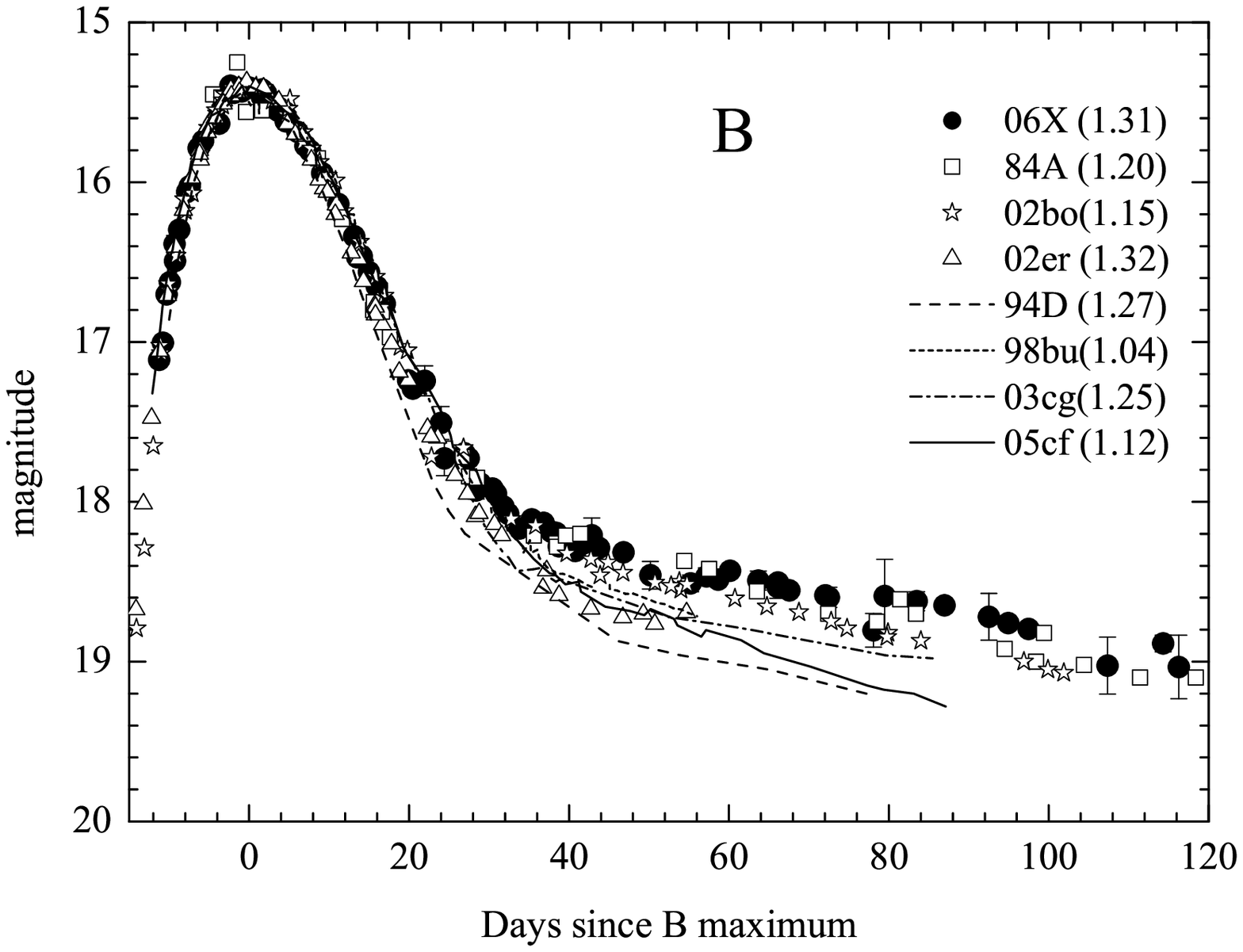}
 \caption{Same as Fig. 3 but for the $B$-band light curve.}
 \label{fig-4}
\end{figure}

\begin{figure}
\figurenum{5} \plotone {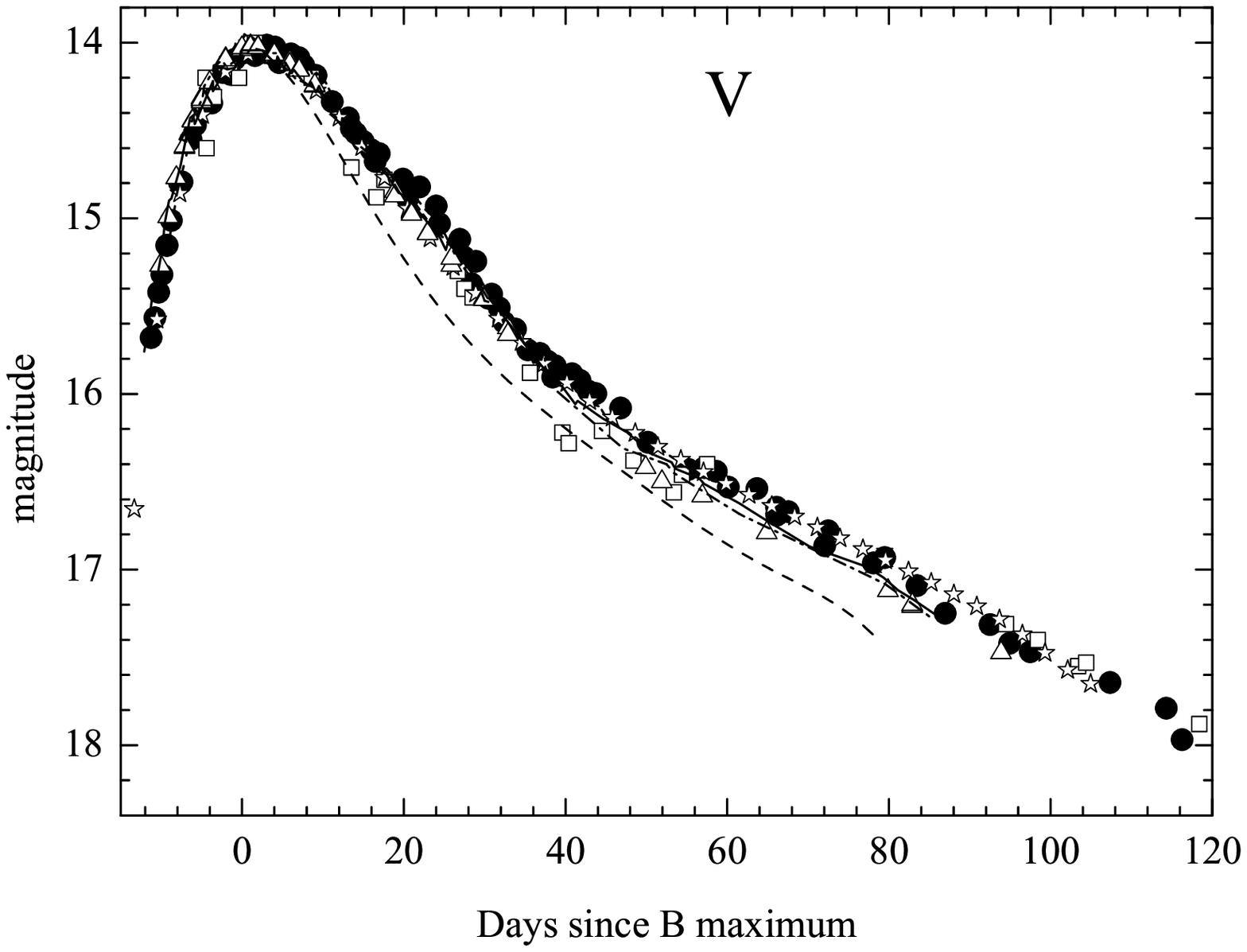}
 \caption{Same as Fig. 3 but for the $V$-band light curve.}
 \label{fig-5}
\end{figure}

\begin{figure}[htbp]
\figurenum{6} \plotone {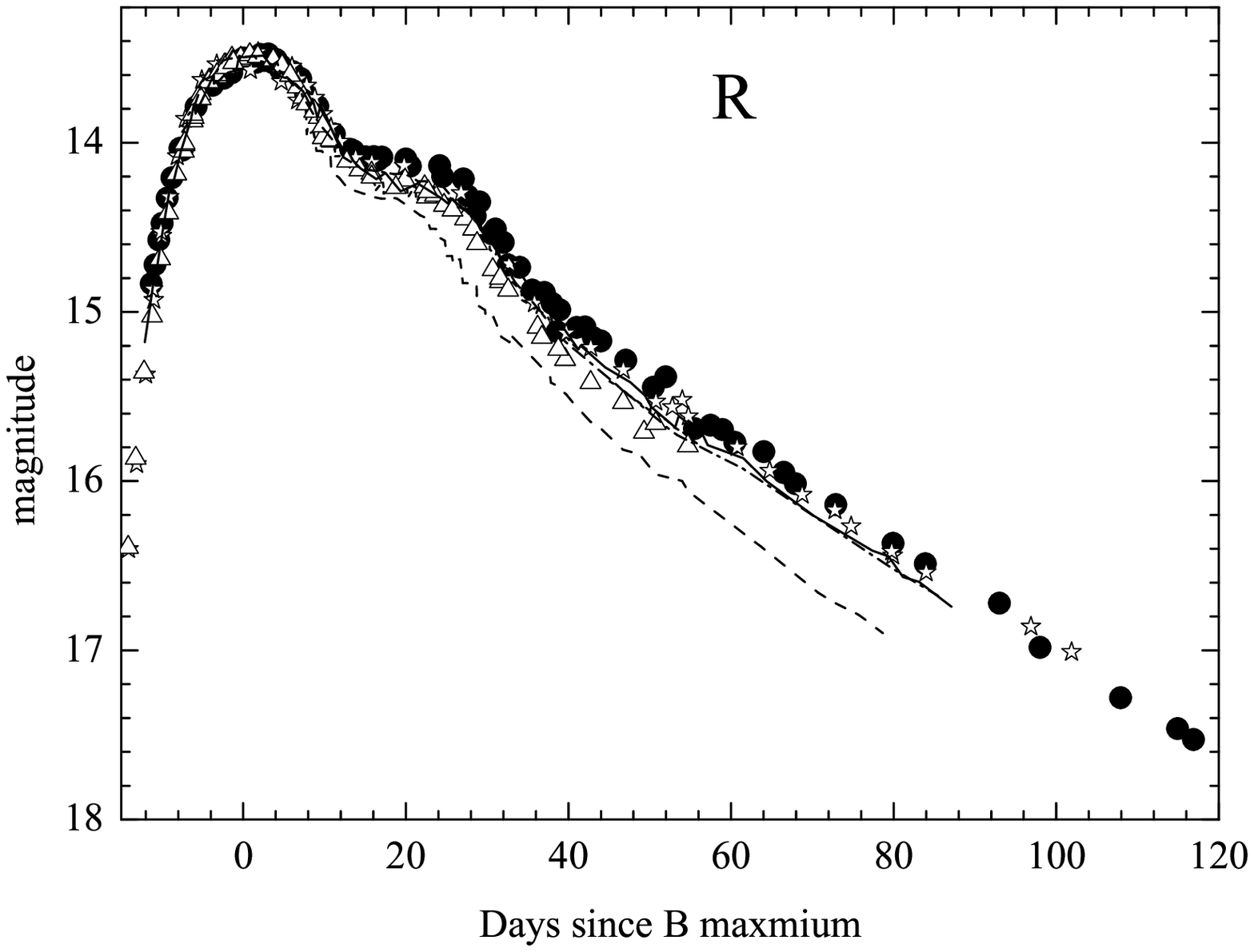}
 \caption{Same as Fig. 3 but for the $R$-band light curve.}
 \label{fig-6}
\end{figure}

\begin{figure}[htbp]
\figurenum{7} \plotone {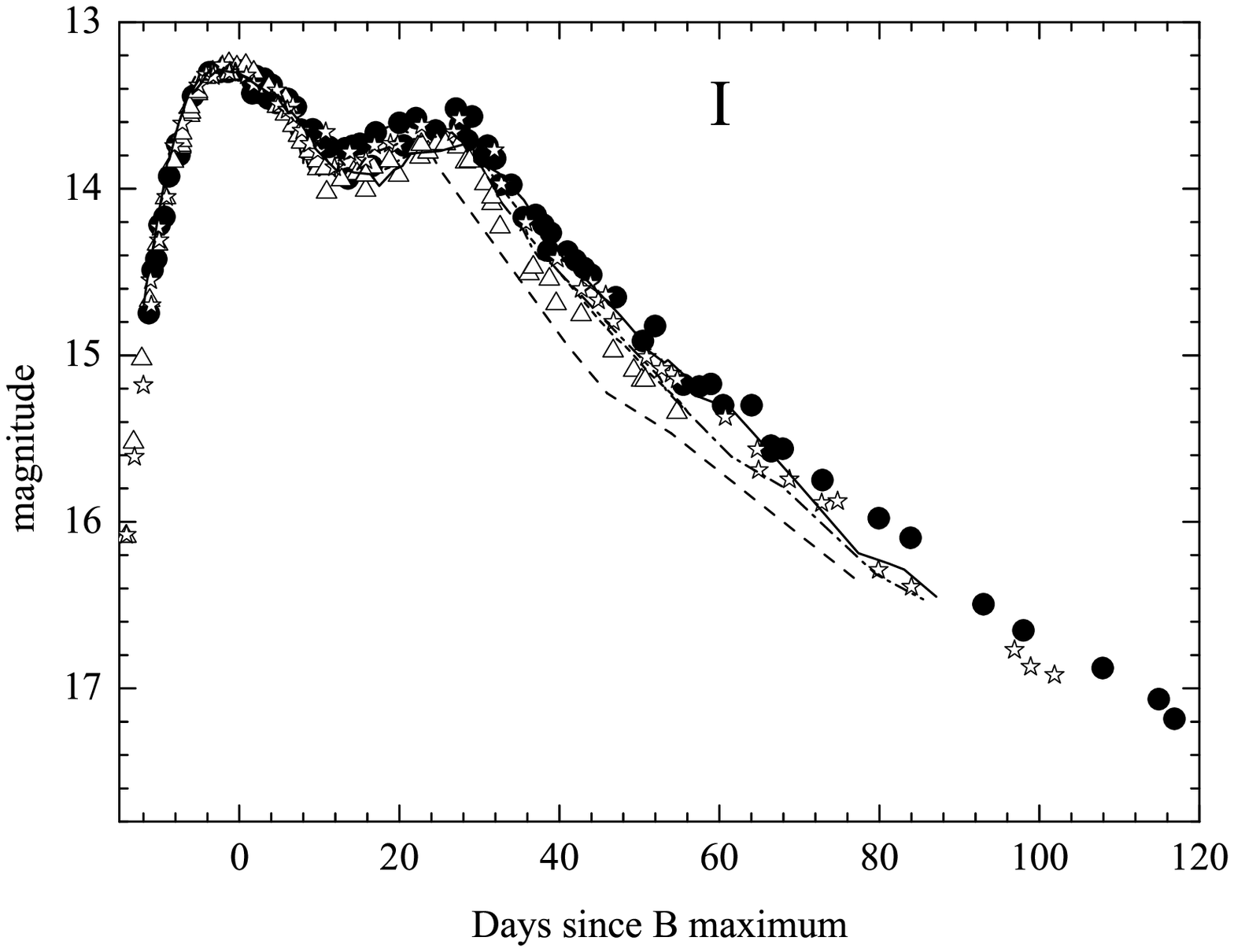}
 \caption{Same as Fig. 3 but for the $I$-band light curve.}
 \label{fig-7}
\end{figure}

\begin{figure}[htbp]
\figurenum{8} \plotone {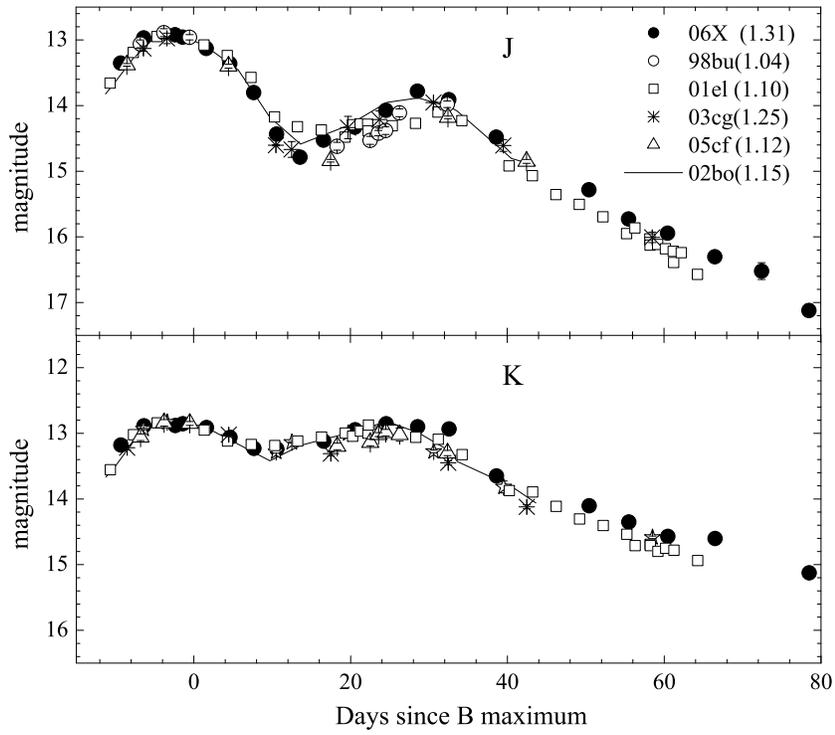}
 \caption{Near-IR $J$-band and $K$-band light curves of SN 2006X
compared with those of SNe 1998bu, 2001el, 2002bo, 2003cg, and
2005cf. The data sources are cited in the text.}
 \label{fig-8}
\end{figure}

\begin{figure}[htbp]
\figurenum{9}
\hspace{-1.5cm}
\includegraphics[angle=0,width=200mm]{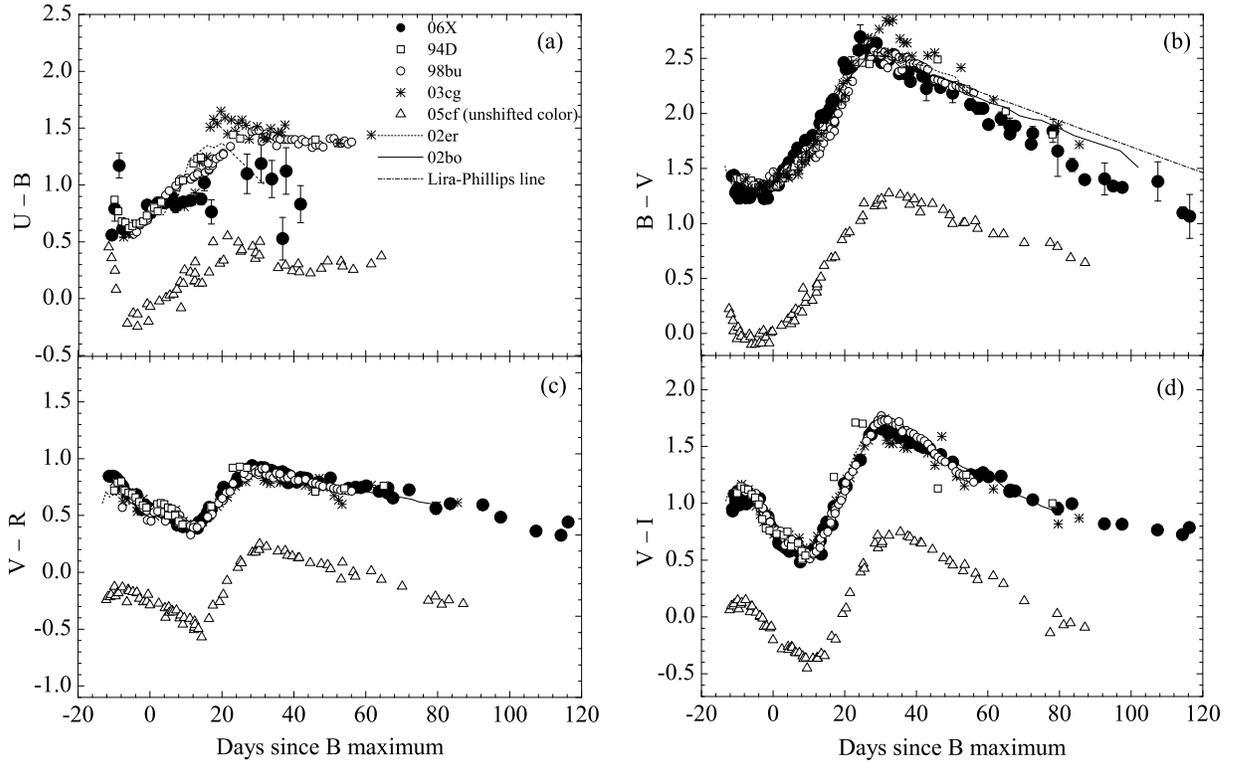}
 \caption{$U - B$, $B - V$, $V - R$, and $V - I$ color curves of
SN 2006X compared with those of SNe 1994D, 1998bu, 2002bo, 2002er,
2003cg, and 2005cf. To show variations in the color evolution
between different SNe Ia, the peak colors of all comparison SNe
(except for SN 2005cf) have been artificially shifted redwards to
match the observed values of SN 2006X at $B$ maximum. The
dash-dotted line shows the Lira-Phillips relation plus a reddening
of $E(B-V) = 1.42$ mag. The data sources are cited in the text.}
 \label{fig-9} \vspace{-0.2cm}
\end{figure}

\begin{figure}
\figurenum{10} \plotone {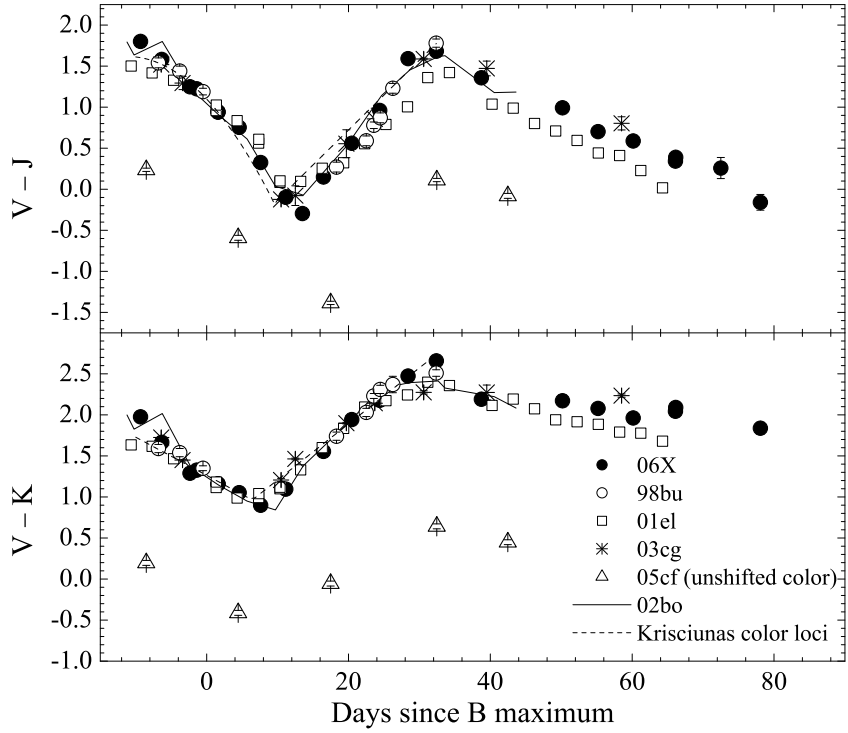} \vspace{1.2cm}
 \caption{The $V - J$ and $V - K$ color curves of SN 2006X, together with
those of SNe 1998bu, 2001el, 2002bo, 2003cg, and 2005cf. The dashed
lines are the loci of the average light curves from Krisciunas et
al. (2000), which are shifted by +1.72 in $V - J$ and +1.95 in $V -
K$ in the ordinate direction to minimize the reduced $\chi^{2}$ of
the fits.}
 \label{fig-10} \vspace{-0.2cm}
\end{figure}

\begin{figure}
\figurenum{11} \plotone {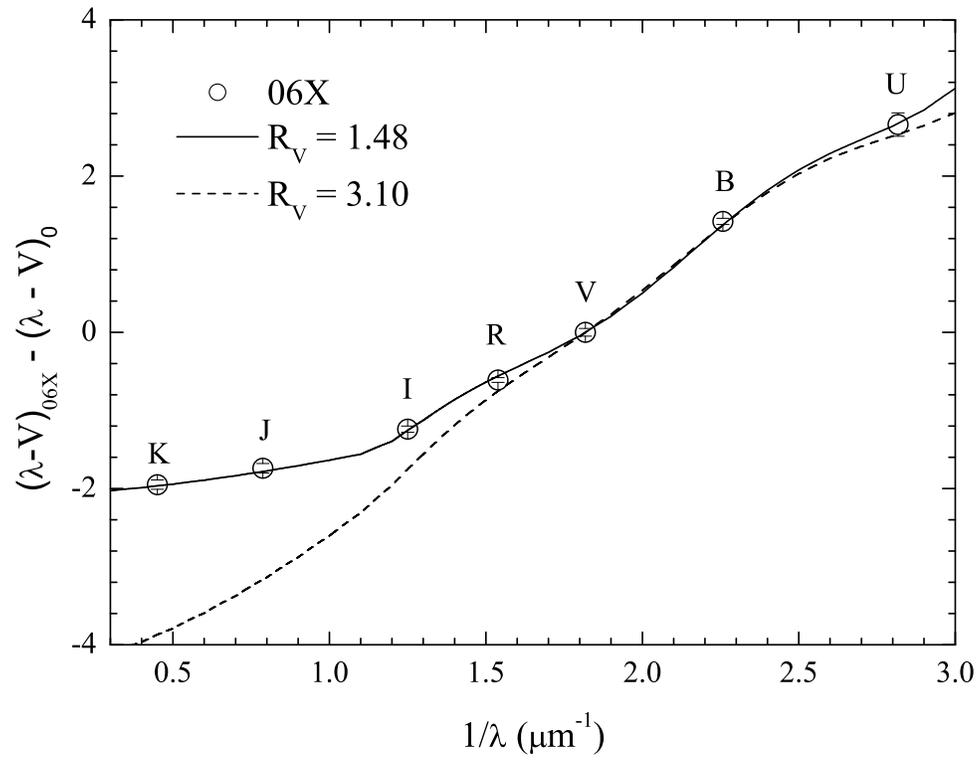}
 \caption{$UVBRIJK$ minus $V$ color differences between SN 2006X and
the unreddened SNe Ia with similar values of $\Delta m_{15}$. The
solid curve shows the best-fit extinction curve following the
analytical model of Cardelli et~al. (1989). The dashed curve
represents the standard extinction curve with $R_{V} = 3.1$.}
 \label{fig-11}
 \end{figure}

\begin{figure}
\figurenum{12} \vspace{-4.5cm} {\plotone{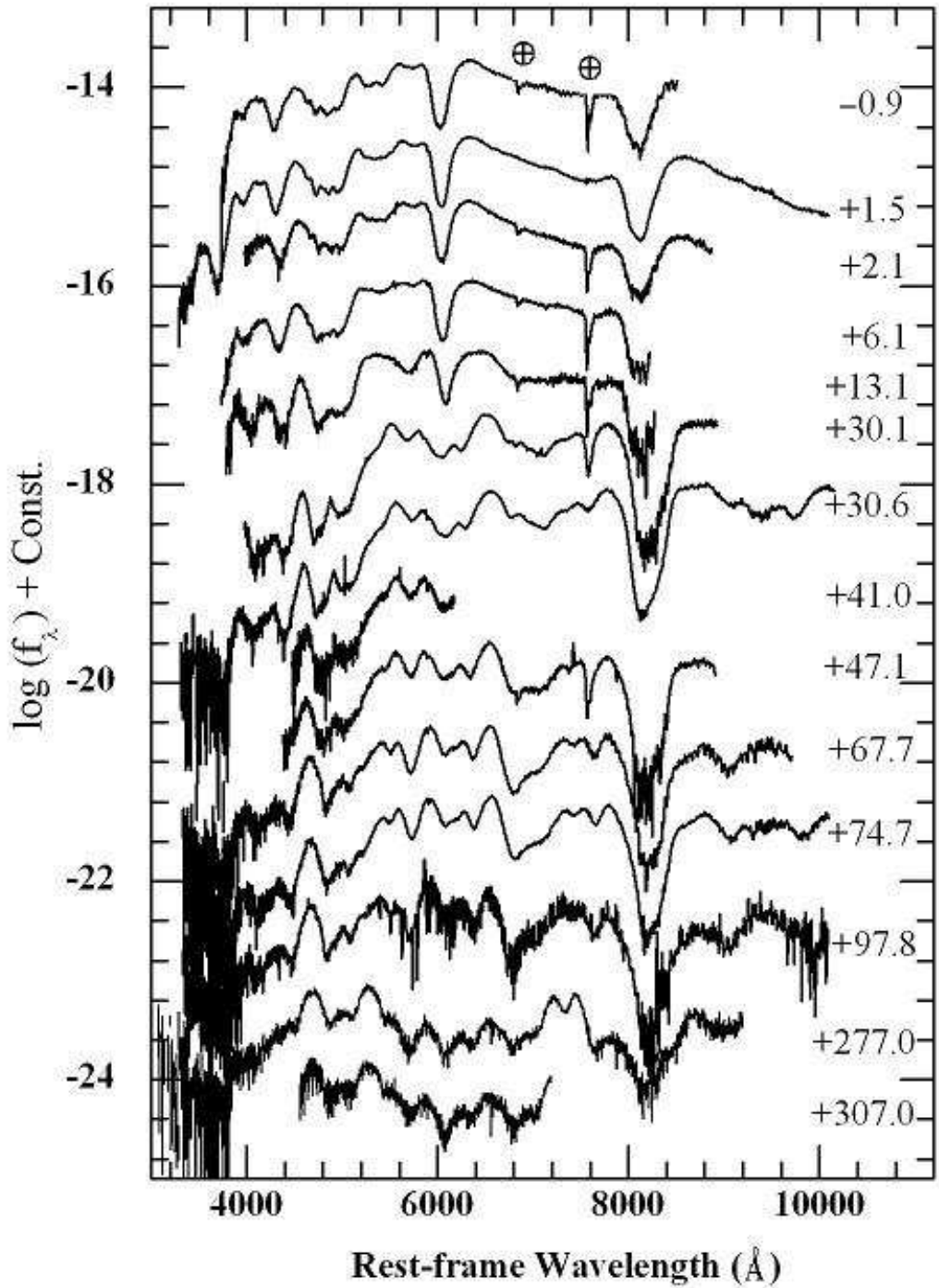}}
\caption{Optical spectral evolution of SN 2006X. The spectra have
been corrected for the redshift of NGC 4321 ($v_{hel}$ = 1567 km
s$^{-1}$) but not reddening, and they have been shifted vertically
by arbitrary amounts for clarity. The numbers on the right-hand side
mark the epochs of the spectra in days after $B$ maximum.
Uncorrected telluric absorption is marked with a plus sign inside a
circle.} \label{fig-12}
\end{figure}

\clearpage
\begin{figure}
\figurenum{13} \hspace{-0.5cm} \plotone{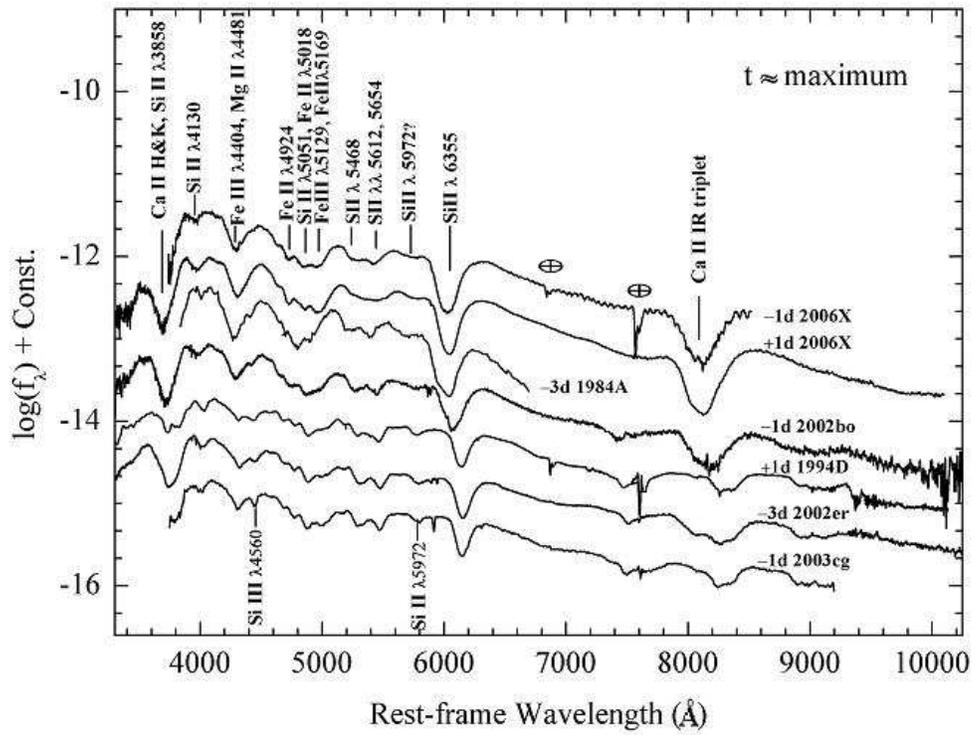} \hspace{-0.5cm}
\caption{The spectrum of SN 2006X near $B$ maximum brightness,
overplotted with comparable-phase spectra of SNe 1984A, 1994D,
2002bo, 2002er, and 2003cg. All spectra shown here and in the
subsequent figures have been corrected for reddening and redshift.
For clarity of display, the spectra were arbitrarily shifted in the
vertical direction.} \label{fig-13}
\end{figure}

\begin{figure}
\figurenum{14} \hspace{-0.5cm} {\plotone{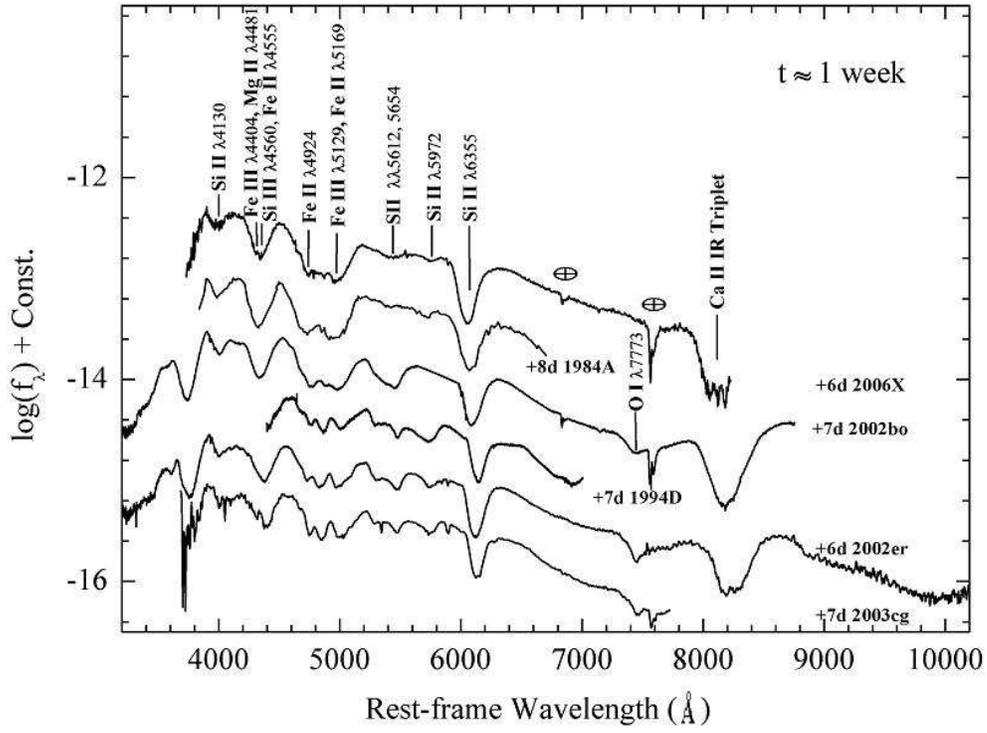}} \hspace{-0.5cm}
\caption{Same as Fig. 13 but for spectra about one week after $B$
maximum.} \label{fig-14}
\end{figure}

\begin{figure}
\figurenum{15} \hspace{-0.5cm} {\plotone{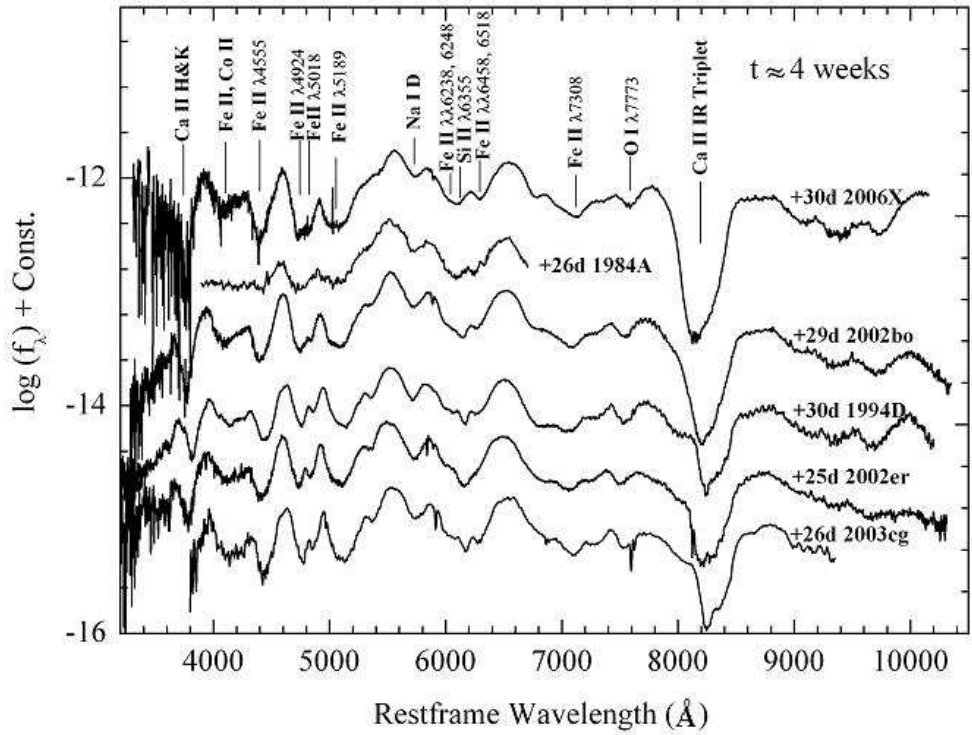}} \vspace{-0.2cm}
\caption{Same as Fig. 13 but for spectra at about 4 weeks past $B$
maximum.} \label{fig-15}
\end{figure}

\begin{figure}
\figurenum{16} \hspace{-0.5cm} {\plotone{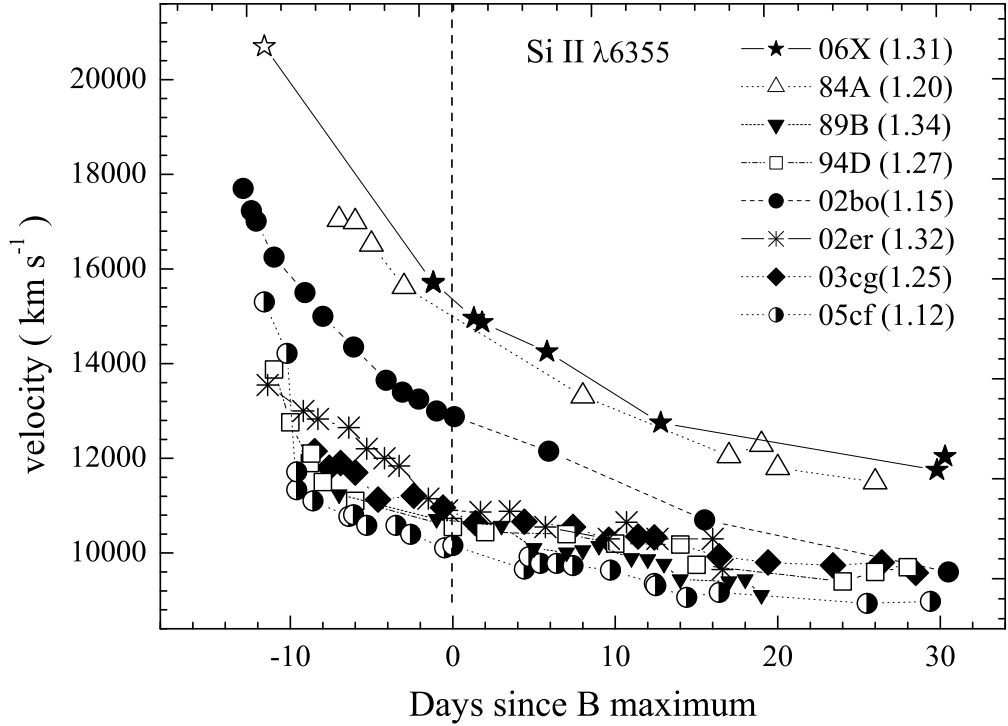}} \hspace{-0.5cm}
\caption{Evolution of the expansion velocity of SN 2006X as measured
from the minimum of Si~II $\lambda$6355, compared with the values of
SNe 1984A taken from Barbon et~al. (1989), 1989B from Barbon et~al.
(1990), 2002bo from Benetti et~al. (2004), 1994D from Patat et~al.
(1996), 2003cg from Elias-Rosa et~al. (2006), 2002er from Pignata
et~al. (2004), and 2005cf from Garavini et~al. (2007). Some of the
data points in the plot are from our unpublished spectral library.
The numbers in parentheses represent the $\Delta m_{15}$ values for
the SNe~Ia.} \label{fig-16}
\end{figure}

\begin{figure}
\figurenum{17} \hspace{-0.5cm} {\plotone{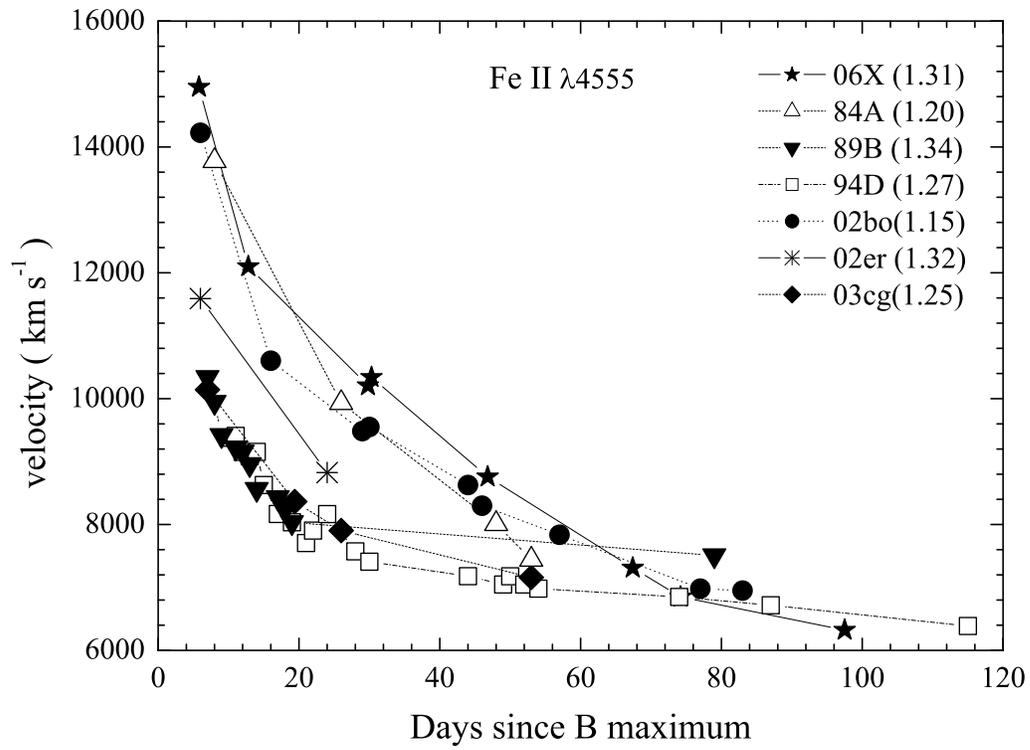}} \hspace{-0.5cm}
\caption{The same as Fig. 16, but for the Fe~II $\lambda$4555 line.}
\label{fig-17}
\end{figure}

\begin{figure}
\figurenum{18} \vspace{-4.5cm} {\plotone{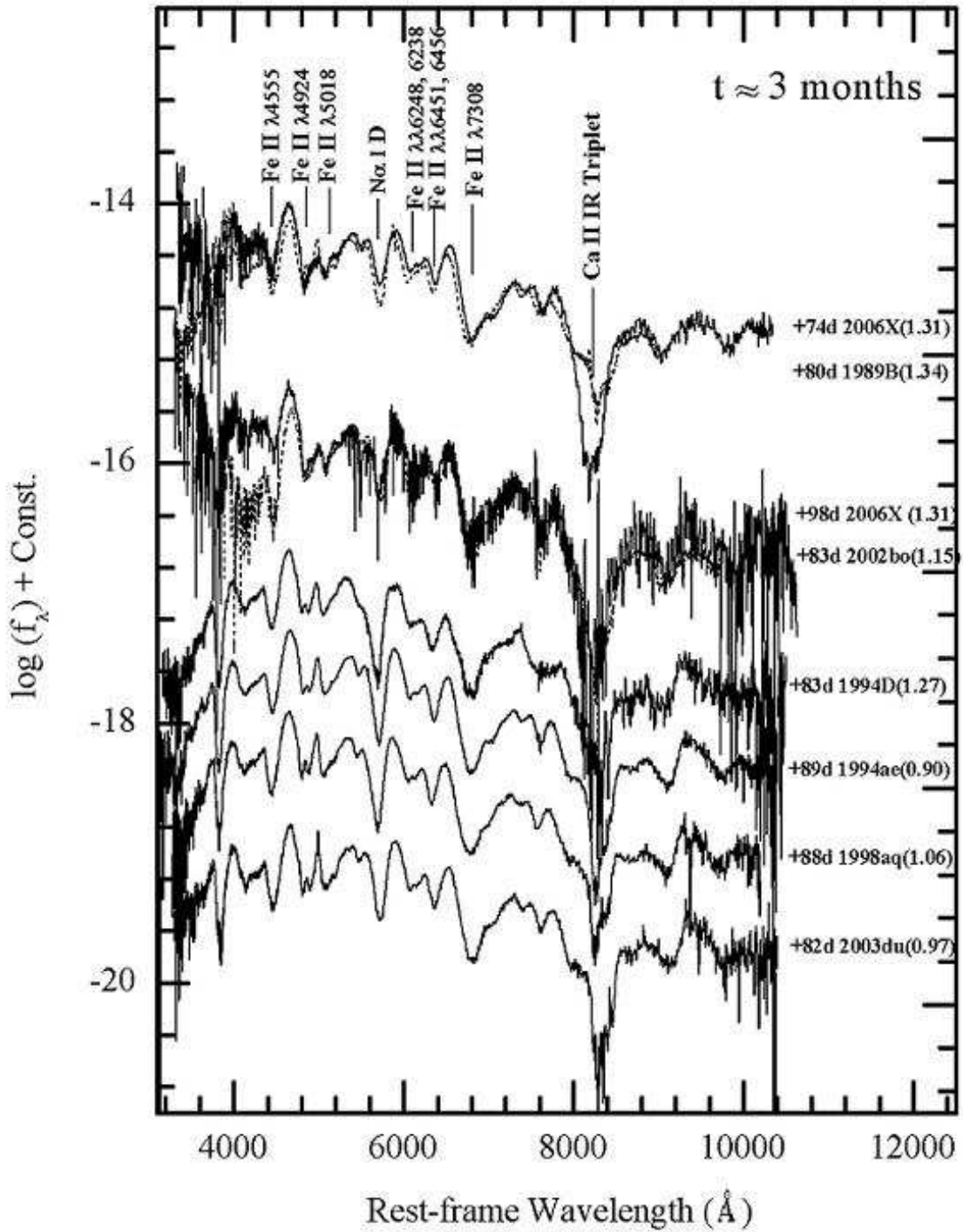}} \hspace{-0.5cm}
\caption{The early-epoch nebular spectra of SN 2006X compared with
those of other SNe~Ia at three months past $B$ maximum. The dashed
spectra overlapping the spectra of SN 2006X at days +75 and +98 are
those of SNe 1989B and 2002bo at similar epochs. The comparison
spectra are from our unpublished spectral library. All the spectra
have been vertically shifted for clarity. The numbers in parentheses
represent the $\Delta m_{15}$ values for the SNe~Ia.} \label{fig-18}
\end{figure}

\begin{figure}
\figurenum{19} \hspace{-0.5cm} {\plotone{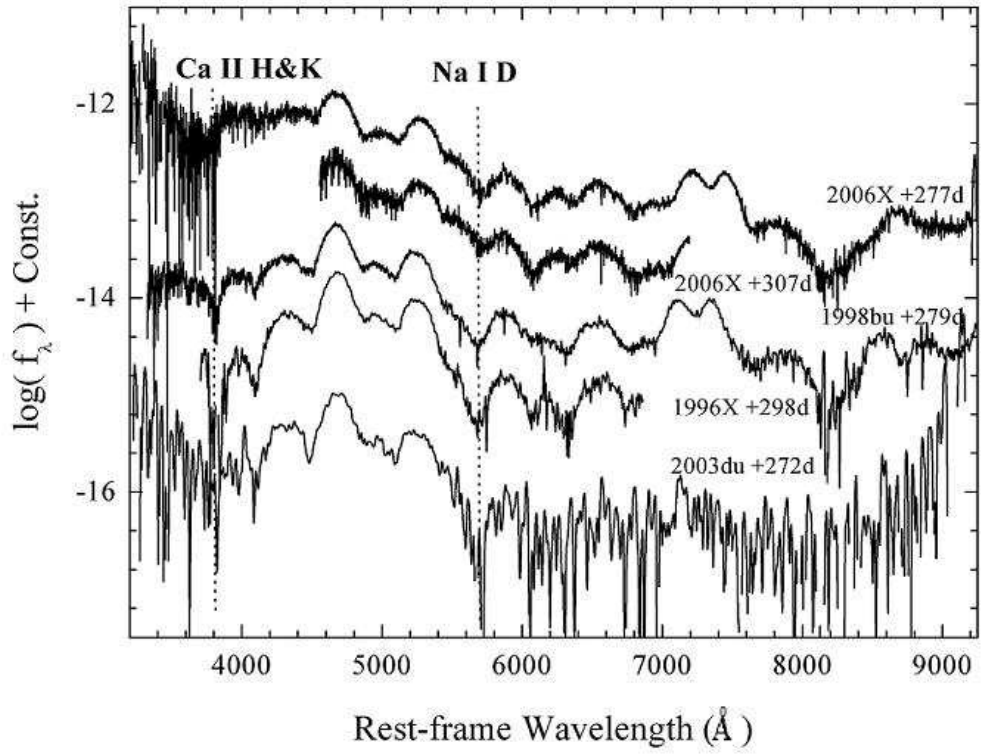}} \hspace{-0.5cm}
\caption{Very late-time nebular spectrum of SN 2006X compared with
those of SNe 1996X, 1998bu, and 2003du. The two vertical dotted
lines refer to the Ca~II $\lambda$3945 and Na~I $\lambda$5892
absorptions.} \label{fig-19}
\end{figure}

\begin{figure}[htbp]
\figurenum{20} \plotone {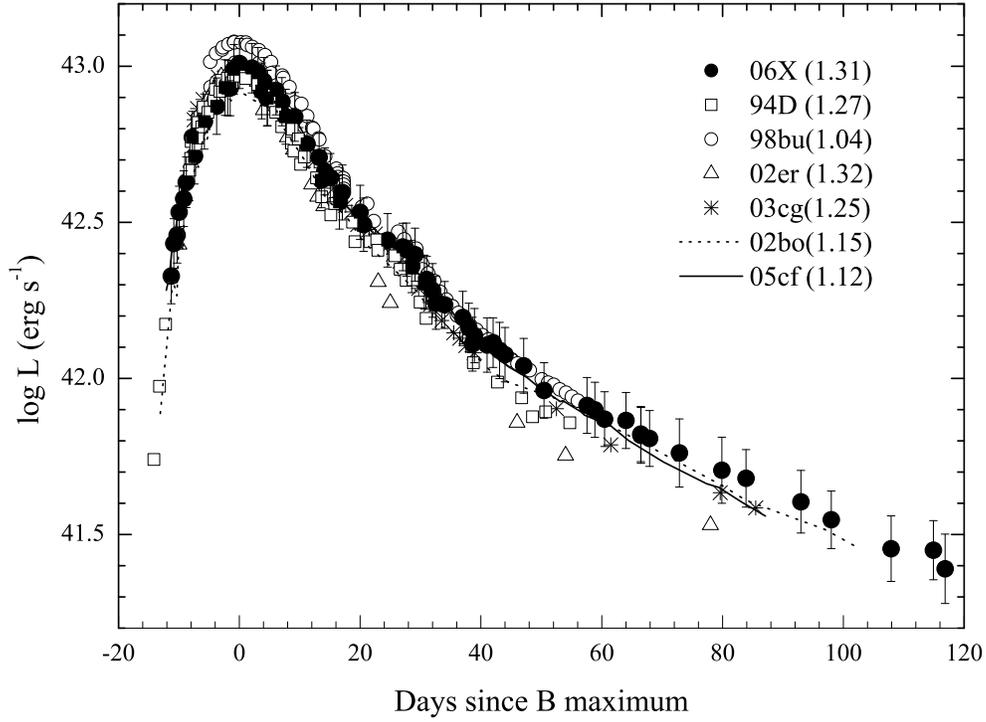}
 \caption{The quasi-bolometric light curve of SN 2006X computed from the
$UBVRIJK$ light curves. Error bars consist of the uncertainties in
the photometry, the distance and absorption corrections. The
quasi-bolometric light curves constructed from SNe 1994D, 1998bu,
2002bo, 2002er, 2003cg, and 2005cf are overplotted for comparison
(see \S 3.1 for the data sources). The numbers in parentheses
represent the $\Delta m_{15}$ values for the SNe~Ia.}
 \label{fig-20}
\end{figure}


\begin{thebibliography}{}
\bibitem [Altavilla et al.(2004)]{alta04}
   Altavilla, G., et~al. 2004, MNRAS, 349, 1344
\bibitem [Arnett 1982]{arne82} Arnett, W. D., 1982, \apj, 253, 785
\bibitem [Arnett 1985]{arne85} Arnett, W. D., Branch D., \& Wheeler J. C., 1985,
Nature, 314, 337
\bibitem [Astier et al. (2006)]{ast06} Astier, P., et al. 2006, \aap, 447, 31
\bibitem [Barbon  (1989)]{bar89} Barbon, B., Rosino L., \& Iijima, T. 1989, \aap,
220, 83
\bibitem [Barbon et~al. (1990)]{bar90} Barbon, B., Benetti, S.,
Cappellaro, E., Rosino, L., Turrato, M. 1990, \aap, 237, 79
\bibitem [Benetti et~al.(2005)]{ben05} Benetti, S., et~al. 2005, \apj, 623, 1011
\bibitem [Bessell 1990]{besel90} Bessell, M. S. 1990, \pasp, 102, 1181
\bibitem [Branch 2005]{bran05}Branch D., et~al. 2005, \pasp, 117, 545
\bibitem [Branch 2006]{bran06}Branch D., et~al. 2006, \pasp, 118, 560
\bibitem [Cardelli et~al. (1989)]{card89} Cardelli, J. A., Clayton, G. C., \&
Mathis, J. S. 1989, \apj, 345, 245
\bibitem [Cappellaro et~al. (2001)]{capp01} Cappellaro, E., et~al. 2001, \apj, 549, L215
\bibitem [Cousins 1981]{cous81} Cousin, A. W. J. 1981, South African Astron. Obs. Circ. 6, 4
\bibitem [Eisenstein et~al. (2005)]{eisen05} Eisenstein, D. J., et~al. 2005, \apj, 633, 560
\bibitem [Elias-Rosa et~al.(2006)]{elias06} Elias-Rosa, N., et~al. 2006, \mnras,
369, 1880
\bibitem [Elmhamdi et~al. (2003)]{elm03} Elmhamdi, A., Chugai, N. N., \& Danziger,
I. J. 2003, \aap, 404, 1077
\bibitem [Ferarrese et~al. (1997)]{Fera97} Ferarrese, L., et~al. 1997, \apj, 475, 853
\bibitem [Filippenko (1982)]{fili82} Filippenko, A. V. 1982, \pasp, 94, 715
\bibitem [Filippenko (1997)]{fili97}Filippenko, A. V. 1997, \araa, 35, 309
\bibitem [Filippenko (2005)]{fili05}Filippenko, A. V. 2005, in White
  Dwarfs: Cosmological and Galactic Probes, ed. E. M. Sion, S. Vennes,
  \& H. L. Shipman (Dordrecht: Springer), 97
\bibitem [Filippenko et~al. (2001)]{fili01} Filippenko, A. V., Li, W., Treffers, R.
R., \& Modjaz, M. 2001, in Small Telescope Astronomy on Global
Scales, ed. B. Paczy\'{n}ski, W.-P. Chen, \& C. Lemme (San
Francisco: ASP), p. 121
\bibitem [Foley et~al. (2003)]{foley03} Foley, R. J., et~al. 2003, \pasp, 115, 1220
\bibitem [Freedman et~al. (2001)]{fre01} Freedman, W. L., et~al. 2001, \apj, 553, 47
\bibitem [Gerardy et~al. (2004)]{ger04} Gerardy, C., et~al.  2004, \apj, 607, 391
\bibitem [Goldhaber (1998)]{gold98} Goldhaber, G. 1998, BAAS, 193, 47.13
\bibitem [Hamuy et~al. (1996)]{ham96} Hamuy, M., Phillips, M. M., Suntzeff, N.B.,
   Schommer, R.A., Maza, J., \& Aviles, R. 1996, \aj, 112, 2398
\bibitem [Ho, Filippenko, \& Sargent (1997)]{hfs97} Ho, L. C., Filippenko, A. V.,
  \& Sargent, W. L. W. 1997, ApJS, 112, 315
\bibitem [Iben \& Tutukov (1984)]{Ibtu84} Iben I. Jr., \& Tutukov A. 1984, \apjs,
55, 335
\bibitem [Jeffery \& Branch (1990)]{JB90} Jeffery, D. J., \& Branch, D. 1990
in Supernovae, Jerusalem Winter School for Theoretical Physics, ed.
Wheeler J. C., Piran T., \& Weinberg, S. (Singapore: World
Scientific), p149
\bibitem [Jha (1999)]{jha99} Jha, S., et~al. 1999, \apjs, 125, 73
\bibitem [Jha (2007)]{jha07} Jha, S., Riess A. G., \& Kirshner, R. P.  2007, \apj,
659, 122
\bibitem [Johnson (1966)]{john66} Johnson, H. L., Iriarte, B., Mitchell, R. I.,
Wisniewskj, W. Z. 1966, Commun. Lunar Planet. Lab., 4, 99
\bibitem [Kasen (2006)]{kasen06} Kasen, D. 2006, \apj, 649, 939
\bibitem [Kasen \& Woosley (2007)]{kasen07} Kasen, D., \& Woosley, S. E. 2007, \apj, 656, 661
\bibitem [Knop et~al.(2003)]{knop03} Knop, R. A., et~al. 2003, ApJ, 598, 102
\bibitem [Krisciunas et~al. (2006)]{kris06} Krisciunas, K., Prieto, J. L.,
Garnavich, P. M., Riley, Jessica-Lynn, G., Rest, A., Stubbs, C., \&
McMillan, R. 2006, \aj, 131, 1639
\bibitem [Krisciunas et~al. (2000)]{kris00} Krisciunas, K., et~al. 2000, \apj, 539, 658
\bibitem [Krisciunas et~al. (2003)]{kris03} Krisciunas, K., et~al. 2003, \aj, 125, 166
\bibitem [Krisciunas et~al. (2007)]{kris07} Krisciunas, K., et~al. 2007, \aj, 133, 58
\bibitem [Lentz et~al. (2000)]{lentz00} Lentz, E. J., Baron, E.,
Branch, D., Hauschildt, P. H., \& Nugent, P. E. 2000, \apj, 530, 966
\bibitem [Lira 1995]{lir95} Lira, P. 1995, Masters thesis, University of Chile
\bibitem [Li, et~al. (2001)]{li01} Li, W., et~al. 2001, \pasp, 113, 1178
\bibitem [Mazalli et~al. (1993)]{maza93} Mazzali, P. A., Lucy, L.
B., Danziger, I. J., Couiffes C., Cappellaro, E., Turatto, M., 1993,
\aap, 269, 423
\bibitem [Milne et~al. (2001)] {milne01} Milne, P. A., The, L. S., \& Leising, M. D.
2001, \apj, 559, 1019
\bibitem [Miller \& Stone (1993)]{mill93} Miller, J. S., \& Stone, R. P. S. 1993,
Lick Obs. Tech. Rep. No. 66
\bibitem [Nomoto et~al. (1997)]{nom97} Nomoto, K.,Iwamoto, K., \& Kishimoto, N.
1997, Science, 276, 1378
\bibitem [Nugent et~al. 1995]{nug95} Nugent, P., et~al. 1995, \apj, 455, L147
\bibitem [Oke et~al. (1995)]{Oke95} Oke, J. B., et~al. 1995, \pasp, 107, 375
\bibitem [Parodi et~al. (2000)]{par00} Parodi, B. R., et~al. 2000, \apj, 540, 634
\bibitem [Pastorello et~al.(2007)]{past07} Pastorello, A., et~al. 2007, \mnras, 376,1301
\bibitem [Patat et~al. (1996)]{pat96} Patat,  F., Benetti, S., Cappellaro, E.,
Danziger, I. J., Della Valle, M., Mazzali, P. A., \& Turatto, M.
1996, \mnras, 278, 111
\bibitem [Patat et~al. (2006)]{pat06} Patat, F. 2006, \mnras, 369, 1949
\bibitem [Patat et al. (2007)]{pat07} Patat F. 2007, Science, 317, 924 (P07)
\bibitem [Perlmutter et~al. (1999)]{perl99} Perlmutter, S., et~al. 1999, \apj, 517, 565
\bibitem [Persson et~al. (1998)]{per98} Persson, S. E., Murphy, D. C., Krzeminski,
W., Roth, M., \& Rieke, M. J. 1998, \aj, 116, 2475
\bibitem [Phillips (1993)]{phi93} Phillips, M. M. 1993, \apj, 413, L105
\bibitem [Phillips (1999)]{phi99} Phillips, M. M., et~al. 1999, \aj, 118, 1766 (P99)
\bibitem [Pignata et~al.(2004)]{pig04} Pignata, G., et~al. 2004, \mnras, 355, 178
\bibitem [Prieto et~al. (2006)]{pri06} Prieto, J. L., Rest, A., \& Suntzeff, N. B.
2006, \apj, 647, 501
\bibitem [Quimby et~al. (2006)]{quim06} Quimby, R., Brown, P., \& Gerardy, C. 2006,
CBET, 393
\bibitem [Reindl et~al.(2005)]{Rei05} Reindl, B., Tammann, G. A., Sandage, A., \&
Saha, A. 2005, \apj, 624, 532
\bibitem [Richmond et~al. (1995)]{Rich95} Richmond, M. W., et~al. 1995, \aj, 109, 2121
\bibitem [Riess et~al. (1996)]{rie96} Riess, A. G., Press, W. H., \& Kirshner, R. P.
1996, \apj, 473, 588
\bibitem [Riess et~al. (1998)]{rie98} Riess, A. G., et~al. 1998, \aj, 116, 1009
\bibitem [Riess et~al. (2005)]{Rie05} Riess, A. G., et~al. 2005, \apj, 627, 579
\bibitem [Riess et~al. (2007)]{Rie07} Riess, A. G., et~al. 2007, \apj, 659, 98
\bibitem [Salvo et al. (2001)]{sal01} Salvo M. E., et~al. 2001, \mnras, 321, 254
\bibitem [Schlegel et~al. (1998)]{sfd98} Schlegel, D. J., Finkbeiner, D. P., \&
Davis, M. 1998, \apj, 500, 525
\bibitem [Spergel et~al. (2003)]{sper03} Spergel, D. N., et~al. 2003, \apjs, 148, 175
\bibitem [Spergel et~al. (2007)]{sper07} Spergel, D. N., et~al. 2007, \apjs, 170, 377
\bibitem [Spyromilio et~al.(2004)]{spy04} Spyromilio, J., Gilmozzi, R., Sollerman,
J., Leibundgut, B., Fransson, C., \& Cuby, J. G. 2004, \aap, 426,547
\bibitem [Stetson 1987]{stet87} Stetson, P. B. 1987, \pasp, 99, 191
\bibitem [Stritzinger et~al. (2002)]{striz02} Stritzinger, M., et~al. 2002, \aj,
124, 2100
\bibitem [Stritzinger \& Leibundgut (2005)]{stri05} Stritzinger, M., \& Leibundgut,
B., 2005, \aap, 431, 423
\bibitem [Strovink (2007)]{strov07} Srovink, M. 2007, ApJ, in press (astro-ph/07050726)
\bibitem [Suntzeff 1996]{suntz06} Suntzeff, N. B. 1996, in Supernova and Supernova
Remnants, ed. R. McCray \& Z. R. Wang (Cambridge: Cambridge Univ.
Press), p. 41
\bibitem [Suzuki \&  Migliardi]{SM06} Suzuki S., \& Migliardi M. 2006, IAU Circ. 8667
\bibitem [Tripp (1997)]{trip97} Tripp, R. 1997, \aap, 325, 871
\bibitem [Tonry et~al. (2003)]{ton03} Tonry, J. L., et~al. 2003, \apj, 594, 1
\bibitem [Wang L (2005)]{wan05} Wang, L. 2005, \apj, 635, L33
\bibitem [Wang L (2006a)]{wan06a} Wang, L., et~al. 2006a, \apj, 641, 50
\bibitem [Wang L (2006b)]{wan06b} Wang, L., et~al. 2006b, \apj, 653, 490
\bibitem [Wang L (2006c)]{wan06c} Wang, L., Baade, D., Patat, F., \& Wheeler, J. C.
2006c, CBET, 396
\bibitem [Wang et~al. (2005)]{wang05} Wang, X., Wang, L., Zhou, X., Lou, Y., \& Li,
Z. 2005, \apj, 620, L87
\bibitem [Wang et~al. (2006)]{wang06} Wang, X., et~al. 2006, \apj, 645, 488
\bibitem [Wang et~al. (2007)]{wang07} Wang, X., et~al. 2007, \apj, submitted.
\bibitem [Webbink (1984)]{web84} Webbink, R. F. 1984, \apj, 277, 355
\bibitem [Wood-Vasey et~al. (2007)]{wood07} Wood-Vasey, W. M., et~al. 2007, \apj, in
press (astro-ph/0701041)

\end{thebibliography}
\end{document}